\RequirePackage[2018-12-01]{latexrelease}
\documentclass{rQUF2e}

\newif\ifarxiv
\arxivtrue

\usepackage{graphicx}
\usepackage{booktabs}
\usepackage{algorithm}
\usepackage{algpseudocode}

\theoremstyle{plain}
\newtheorem{theorem}{Theorem}

\newtheorem{lemma}{Lemma}
\newtheorem{proposition}{Proposition}

\theoremstyle{definition}
\newtheorem{assumption}{Assumption}

\theoremstyle{remark}
\newtheorem{remark}{Remark}

\newcommand{\1}{\mathbf{1}}
\newcommand{\KL}{D_{\mathrm{KL}}}
\newcommand{\Safe}{\textsc{Safe}}
\newcommand{\Follow}{\textsc{Follow}}
\newcommand{\Reverse}{\textsc{Reverse}}

\newcommand{\principle}[1]{%
  {\setlength{\fboxsep}{7pt}\fbox{\parbox{0.84\linewidth}{\centering #1}}}}

\ifarxiv
\makeatletter
\def\ps@plain{\let\@mkboth\@gobbletwo
  \let\@oddhead\@empty\let\@evenhead\@empty
  \let\@oddfoot\@empty\let\@evenfoot\@empty}
\def\receivedfont{\vskip-18pt\let\rm\@gobble\let\@received\@empty}
\def\draftnote{}
\makeatother
\fi

\begin{document}

\title{Safe Portfolio Learning from Untrusted Predictions: Signed Trust, Abstention, and Cross-Universe Polarity Failure}

\author{Amirhossein MohammadpourAzari$\dag$, Mohammad Zarei$\dag$ and MohammadAmin Fazli$^{\ast}$$\dag$\thanks{$^\ast$Corresponding author. Email: fazli@sharif.edu}\\
\affil{$\dag$Department of Computer Science and Engineering, Sharif University of Technology, Tehran, 11365-11155, Iran}
\received{\today} }

\maketitle

\begin{abstract}
Portfolio use of machine predictions must often proceed while the predictions' economic direction, reliability, and timing remain untrusted, and while feedback from overlapping multi-period positions arrives with delay. A controller chooses among Safe, Follow, and Reverse portfolio actions, so signal orientation is learned rather than assumed. A share-adaptive master updates action probabilities from matured Safe-relative rewards, an evidence gate can abstain by transferring unsupported directional probability to Safe, a delayed collateral budget reserves worst-case loss capacity before exposure is funded, and a protected split account caps the capital adaptive decisions may touch. The architecture carries a dual certificate, tracking regret against switching action sequences under delay and one-sided intervention, together with a pathwise collateral safety bound that holds when predictions fail. On a 24-stock development universe a conservative prediction-assisted overlay stays near equal-weight performance at a fraction of the turnover. On a frozen 82-stock holdout, the pre-specified Reverse orientation
fails to outperform Equal Weight; a separately labeled post-holdout
diagnostic shows that permanently flipping the sign to Follow also
does not recover a return edge. Under the post-holdout controller, protected wealth remains within
15.9 basis points of the Safe benchmark throughout. The contribution
is safe adaptive use of untrusted predictions, not demonstrated alpha.
\end{abstract}

\begin{keywords}
Portfolio selection; Online learning; Regret bounds; Transaction costs; Abstention; Distribution shift
\end{keywords}

\section{Introduction}\label{sec:introduction}

Machine predictions for portfolios are now abundant, and the machines disagree. Large language models asked identical investment questions return materially different advice depending on which model answers and how it is prompted, with model-specific default risk tolerances that shift under assigned personas \citep{ChoBaeKim2026RiskProfiles}. Whether such advice carries information at all has to be established against randomized selection rather than assumed \citep{KoLee2023ChatGPTPortfolio}, and when model forecasts steer real allocation decisions, outcomes vary substantially across model families and usage frequency \citep{GuEtAl2024LLMMargin}. Untrustedness is therefore the operating condition of prediction-assisted investing, not an edge case to be assumed away.

The condition is worse than noisy advice, because the economic direction of a signal need not survive a change of setting. Sentiment-fused allocation systems report gains on short out-of-sample windows \citep{LiuEtAl2024NSTD}, and even studies that fuse several sentiment sources into trading agents leave the translation from predicted signal to trading position unresolved \citep{SattarEtAl2025RMSDRL}, so nothing in the reported evidence establishes that a signal profitable under one orientation keeps that orientation elsewhere. This paper contributes a sharp instance of the problem from its own evaluation. A directional specification developed on one equity universe, with its sign fixed in advance, failed to transfer to a disjoint universe, and the failure was two-sided, with neither the frozen orientation nor its permanent opposite beating an equal-weight benchmark. The polarity of a useful prediction turned out to be the unstable quantity.

Portfolio practice is poorly equipped for that instability because it treats signal direction as given. Since mean-variance analysis, allocation has been formalized as a tradeoff whose inputs are presumed reliable \citep{Markowitz1952}, and the online selection literature splits into Follow-the-Winner and Follow-the-Loser families, a dichotomy that presumes the strategy must always express some directional view \citep{LiHoi2014OLPSSurvey}. Forecast accuracy offers no escape, since minimizing forecast error carries no guarantee about the sign of a return \citep{Moody1998}. Direction is consequently not a nuisance parameter that better estimation resolves. It is itself a quantity to learn, under the possibility that at times it should not be acted on at all.

The research question follows. When a prediction's direction, reliability, and timing are all untrusted, what separates producing advice from earning permission to risk capital on it? Four decisions, usually fused, need separating. Which directional actions exist at all, which currently appears economically preferable, whether matured evidence suffices to authorize it, and how much exposure outstanding loss capacity can finance, all under feedback that arrives with a multi-period delay while several decisions remain unresolved.

The answer is a controller built from that separation. Each decision date offers three implementable candidate portfolios, a prediction-independent \Safe{} anchor, a \Follow{} overlay using the signal's stated direction, and a \Reverse{} overlay using its opposite, so orientation becomes a choice rather than an assumption. A share-adaptive master learns a probability over the three actions from matured Safe-relative rewards, hedging over an entire grid of switching intensities so that no belief about polarity persistence is baked in. An evidence gate can veto unsupported directional mass and transfer it to \Safe{}, which is abstention implemented as a fully invested action rather than a refusal to hold assets. A delayed collateral budget reserves worst-case loss capacity for every unresolved position before new exposure is funded. Protected split-account execution then caps, at account level, the fraction of capital the adaptive layers may touch at all.

This architecture admits a dual guarantee, and the duality is the point. A tracking certificate bounds the executed probability path against arbitrary switching sequences of the three actions, paying explicit, named penalties for delayed information and for one-sided safety intervention. A collateral certificate independently bounds cumulative Safe-relative loss pathwise, from capacity reserved in advance, and it holds when predictions fail. The two certificates hold simultaneously, so realized performance is bounded by whichever is stronger on the realized path, without any assumption that either orientation of the advice is correct.

The evidence is organized to earn its claims rather than to flatter them. A development universe of 24 stocks motivated the overlay, showing near-benchmark performance at an order of magnitude less turnover and no return edge. An 82-stock holdout, frozen before inspection, then failed to confirm
the pre-specified Reverse transfer specification. A separately labeled
post-holdout diagnostic showed that permanently flipping the same signal
to Follow also remained below Equal Weight. The negative frozen result
is preserved as the confirmatory outcome rather than repaired after the
fact, a discipline the backtest-overfitting evidence makes non-optional
\citep{BaileyEtAl2016PBO} and whose absence is endemic across
reinforcement-learning finance
\citep{PippasLudvigTurkay2025RLSurvey}. Everything examined after the
frozen verdict is labeled exploratory and stays that way.

Four contributions result. A signed formulation of prediction-assisted allocation in which orientation and abstention are actions rather than assumptions. A safety architecture, evidence gate, delayed collateral budget, and protected split account, whose interventions are one-sided and exactly priced. A dual certificate combining switching-comparator tracking under delay with pathwise collateral safety and a datewise wealth floor. And an evidence hierarchy that preserves a negative transfer result as a finding about polarity instability. The claim boundary is set accordingly. On the realized path protected wealth stayed within 15.9 basis points of the induced \Safe{} benchmark; the contribution is safe adaptive use of untrusted predictions, not demonstrated alpha.

\subsection{Related literature}\label{sec:related}

Growth-optimal online portfolio selection established that allocation can carry worst-case guarantees without market assumptions. Universal portfolios attain the growth rate of the hindsight-optimal rebalanced portfolio on every bounded sequence \citep{Cover1991UP}, extend to conditioning on side information whether or not it proves informative \citep{CoverOrdentlich1996Side}, and admit cheap multiplicative updates \citep{HelmboldEtAl1998EG} and second-order schemes with logarithmic regret \citep{AgarwalEtAl2006ONS}. Proportional transaction costs can be brought inside the guarantee, making rebalancing cost part of what the online strategy trades off \citep{BlumKalai1999TxnCosts}, a line this paper's drift-relative cost accounting continues. What the entire family leaves unexamined is the comparator's directional commitment. Whether universal, Winner-following, or Loser-following \citep{LiHoi2014OLPSSurvey}, the strategy class fixes how signals map to positions, and no guarantee addresses a signal whose economically useful orientation changes.

Expert tracking supplies the machinery for that gap. Weighted-majority aggregation loses only a logarithmic overhead to the best pool member \citep{LittlestoneWarmuth1994WM}, Hedge generalizes multiplicative weights to decision-theoretic allocation \citep{FreundSchapire1997Hedge}, Fixed-Share tracks comparators that switch experts at a cost logarithmic in the horizon and switch count \citep{HerbsterWarmuth1998Tracking}, and mixing past posteriors lets weight return to previously successful experts \citep{BousquetWarmuth2002Mixing}. The paper's master is Fixed-Share machinery under an outer Hedge aggregation, credited as such. What is new is the action set it runs over, three portfolio actions including the signal's negation and a fallback, and the safety layers that stand between its output and capital.

Delayed feedback has its own reductions, minimax subsampling into non-delayed subproblems \citep{WeinbergerOrdentlich2002Delayed} and black-box constructions that inflate adversarial regret multiplicatively in the delay \citep{JoulaniEtAl2013Delayed}. Both hold for arbitrary delay processes and pay for that generality. The overlapping-cohort structure here is not arbitrary, and the paper's reduction is exact, an additive realized penalty measured from the master's own path, with matured Safe-relative rewards as the delayed currency.

Algorithms with predictions formalize the contract this paper transports to capital allocation. Optimism yields tighter regret when a supplied predictable process helps and worst-case control when it fails \citep{RakhlinSridharan2013Predictable}, naively following a learned oracle can be arbitrarily poor while a designed algorithm improves with oracle accuracy and never exceeds its worst-case bound \citep{LykourisVassilvitskii2018Caching,PurohitEtAl2018MLPredictions}, and the general contract, near-optimal when predictions are good and worst-case when they are not, is now a design principle \citep{MitzenmacherVassilvitskii2020AlgPred}. The dual certificate is that contract for portfolios, with one addition the caching problems do not need, a hard pathwise safety branch priced in reserved collateral rather than competitive ratio.

Abstention began as Chow's rejection rule, declining to decide when the posterior falls below a cost-determined threshold \citep{Chow1970Reject}, matured into selective prediction's risk-coverage tradeoff \citep{ElYanivWiener2010Selective} and jointly learned rejection \citep{CortesEtAl2016Rejection}, and in the online-expert setting buys an improvement from square-root to logarithmic regret \citep{NeuZhivotovskiy2020Abstention}. This paper recasts the abstaining action economically. \Safe{} risks no capital on direction while remaining fully invested, so abstention is not a statistical refusal to predict but a portfolio that declines a directional bet.

Safe and risk-sensitive reinforcement learning constrains policies through modified criteria or guided exploration \citep{GarciaFernandez2015SafeRL}, with constrained policy updates guaranteeing violation bounds in expectation \citep{AchiamEtAl2017CPO}, CVaR-triggered overrides of portfolio policies \citep{WangKu2022}, entropic-risk tuning under distribution shift \citep{EndersHarrisonSchiffer2024}, and coherent CVaR objectives inside actor-critic learning \citep{MishraEtAl2026RiskSensitiveCVaR}. The load-bearing contrast is expectation versus pathwise. Those protections hold on average or in distribution under a model of the environment, whereas the collateral certificate here is an accounting invariant on every realized path, and what the fallback represents, a preference the investor already holds, connects to work on learning risk preferences from observed choices \citep{AlsabahEtAl2019}.

Deep and direct reinforcement learning for portfolios runs from performance-function training \citep{Moody1998} and direct policy learning without a forecasting stage \citep{MoodySaffell2001} through end-to-end weight generation with cost feedback \citep{JiangXuLiang2017}, cost-sensitive rewards with near-log-optimal growth \citep{ZhangEtAl2022}, drawdown-based objectives and constrained signal mappings \citep{AlmahdiYang2017,AlmahdiYang2019,GarciaGaliciaEtAl2019,ParkSimChoi2020,ParkEtAl2024TSRRL}, CVaR-based tail control \citep{CuiEtAl2023CVaR}, and standardized actor-critic comparisons \citep{RezaeiNezamabadiPour2025DRLTaxonomy}. Across the line, the reward signal that trains the policy is trusted as given, risk enters through constraints on magnitude rather than through orientation, and no mechanism asks whether the policy's directional premise still holds. Orientation learning of the kind pursued here has no counterpart in this literature.

LLM-advised allocation is the newest advice source and the least filtered. Language models evolve portfolio heuristics \citep{ChenWangZhang2025EPH}, refine goal-conditioned trading decisions \citep{WangEtAl2025GEMsLLM}, label sector news for rotation \citep{YanEtAl2025DRLSR}, reason over sector context for selection \citep{KimEtAl2026SectorAwareLLM}, drive agentic regime-aware controllers \citep{MantshimuliMwamba2026AgenticLLM}, and validate sentiment inputs to mean-variance construction \citep{Chen2026SentimentMVOCrypto}. Closest in spirit to this paper, advice can enter as a regularizing prior rather than a decision-maker, with guarantees that degrade gracefully as the advice worsens \citep{ZhangEtAl2024LINVIT}. What the line as a whole lacks is a capital-authorization layer, any mechanism between generated advice and funded exposure that answers for direction, evidence, and financing.

Evaluation discipline closes the positioning on the gap itself. Single holdout splits are unreliable for strategy evaluation \citep{BaileyEtAl2016PBO}, methodological weaknesses are endemic across surveyed RL finance \citep{PippasLudvigTurkay2025RLSurvey}, feature configuration alone moves DRL risk outcomes at scale \citep{Ayari2026FeatureConfig}, and an 84-study survey of LLM equity applications contains no category for bounding the cost of incorrect model output \citep{JadhavMirza2025LLMEquityReview}. No existing framework bounds the economic cost of acting on wrong directional advice under delayed feedback. The methodology supplies one.

\section{Methodology}\label{sec:methodology}

The controller answers the four questions with four separable mechanisms. What directional actions exist is settled by a candidate layer that builds three implementable portfolios, one of which ignores predictions entirely. Which action currently appears economically preferable is learned by a share-adaptive signed master from matured rewards. Whether that matured evidence suffices to authorize the learned preference is decided by an evidence gate, and how much authorized exposure outstanding loss capacity can finance is decided by a delayed collateral budget, with a protected split account fixing how much total capital the adaptive layers may touch at all. The separation itself is the design contribution, because prediction quality enters only through the candidate actions, whereas learning, abstention, collateral, and account protection are defined without any assumption that a prediction is correct. The decision cycle as pseudocode and the complete collateral bookkeeping are recorded in Appendix~\ref{app:pseudocode}; the remaining mechanism-by-mechanism specification is given once in the online supplement.

\begin{figure}
\centering
\includegraphics[width=\textwidth]{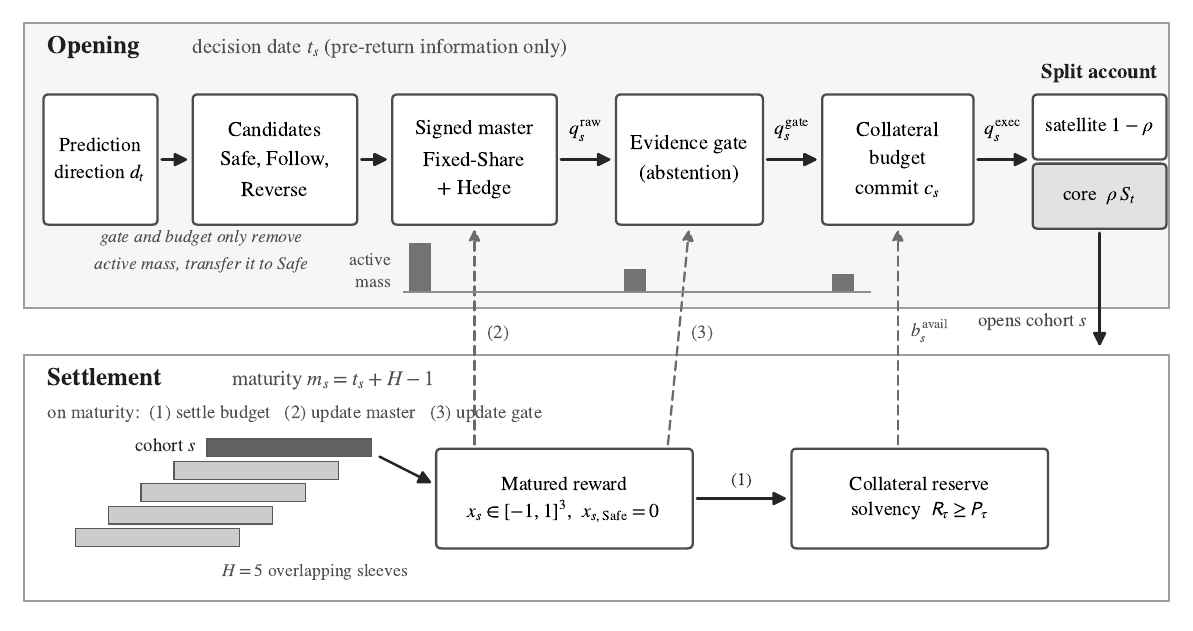}
\caption{The controller on its two clocks. Opening (top): the prediction becomes a centered direction, three candidate portfolios feed the share-adaptive signed master, and the raw probability passes the evidence gate and the collateral budget, both one-sided (falling active-mass bars), before the executed mixture trades in the satellite of the split account whose protected core follows the \Safe{} strategy. Settlement (bottom): cohorts overlap in $H=5$ staggered sleeves, and a maturing cohort's Safe-relative reward settles the collateral reserve, then updates the master, then the gate, in that strict order. No same-date maturity can influence the cohort already opened.}
\label{fig:architecture}
\end{figure}

Figure~\ref{fig:architecture} compresses the composition into its governing invariant. Every mechanism runs on one of two clocks. Decisions occur at opening, on information available before the current return is realized, and produce the executed mixture at the right of the pipeline. Learning states, evidence statistics, and collateral commitments change only at maturity, when a cohort's Safe-relative reward becomes observable. Nothing is permitted to leak between the two clocks, and every definition below respects that ordering.

\subsection{Setting and candidate portfolios}\label{sec:candidate_portfolios}

The market interface is deliberately spare. Over $N$ tradable assets with simple returns $r_t$, a portfolio is a point of the capped long-only simplex, fully invested with single-asset weights limited by $w_{\max}$. Between two decisions the previous target drifts under realized returns, so trading intensity is measured from the drifted holdings rather than from the previous target. A target close to yesterday's target can still require substantial trading when prices have displaced the current holdings, and a distant target can be nearly free when the drift has already moved the portfolio the same way. Turnover is accordingly the full $\ell_1$ displacement from the drifted portfolio, and only that displacement counts as trading,
\begin{equation}
\label{eq:feasible_set}
\begin{gathered}
\Delta_N=\Bigl\{w\in\mathbb{R}^{N}_{+}:\ \1^{\top}w=1,\ w_i\leq w_{\max}\Bigr\},
\qquad
\operatorname{TO}_t=\bigl\|w_t-\widetilde{w}_t\bigr\|_1,
\\
\widetilde{w}_t=\frac{w_{t-1}\odot(1+r_{t-1})}{1+w_{t-1}^{\top}r_{t-1}}.
\end{gathered}
\end{equation}
A proportional coefficient $c\geq0$ converts displacement into cost, so the one-period net return of a target $w_t$ is $w_t^{\top}r_t-c\,\|w_t-\widetilde{w}_t\|_1$. Charging costs against drifted holdings places the design in the cost-aware portfolio-learning line opened by \citet{BlumKalai1999TxnCosts} and continued in displacement-dependent cost modeling \citep{Moroke2026FreeEnergyCrypto}. The value of $c$, like every controller constant below, is fixed in the experimental design.

The prediction-independent anchor is produced by a cost-aware mirror-descent operator applied to a conservative gradient $g_t^{\mathrm{safe}}$ that carries no predictive signal,
\begin{equation}
\label{eq:opmd_operator}
\operatorname{OPMD}\bigl(\widetilde{w}_t,g\bigr)
=\arg\min_{w\in\Delta_N}
\Bigl\{\langle g,w\rangle+\tfrac{1}{\eta_w}\KL\bigl(w\,\|\,\widetilde{w}_t\bigr)+\lambda\bigl\|w-\widetilde{w}_t\bigr\|_1\Bigr\},
\qquad
w_t^{\Safe}=\operatorname{OPMD}\bigl(\widetilde{w}_t,g_t^{\mathrm{safe}}\bigr).
\end{equation}
The construction instantiates the optimistic mirror-descent template of \citet{RakhlinSridharan2013Predictable}. The Kullback--Leibler proximal term with step size $\eta_w$ keeps the update close to the drifted holdings in the geometry of the simplex, while the $\ell_1$ penalty $\lambda$ prices movement directly; the \Safe{} targets used in every experiment are those produced by the archived development pipeline, with $\eta_w=0.10$ and $\lambda$ tied to the cost coefficient. \Safe{} is consequently a fully invested, low-turnover portfolio, not cash and not an abstention from investment. That reading matters for everything that follows, because the safety layers transfer probability to \Safe{}, and the transfer would mean something different if \Safe{} stood outside the market.

The two active candidates are signed overlays of a common direction on the same anchor,
\begin{equation}
\label{eq:follow_reverse}
\begin{gathered}
w_t^{\Follow}=w_t^{\Safe}+\alpha_t^{+}\delta_t d_t,
\qquad
w_t^{\Reverse}=w_t^{\Safe}-\alpha_t^{-}\delta_t d_t,
\\
\alpha_t^{\pm}=\max\bigl\{\alpha\in[0,1]:\ w_t^{\Safe}\pm\alpha\delta_t d_t\in\Delta_N\bigr\}.
\end{gathered}
\end{equation}
The direction $d_t$ is the centered unit-$\ell_1$ image of the cross-sectional prediction, with $\1^{\top}d_t=0$ so the overlay redistributes mass without changing total investment, and $\delta_t$ is the requested overlay magnitude. Both orientations are necessary because a fixed-sign design cannot distinguish a weak forecast from a forecast whose useful orientation has changed. \Follow{} asks whether the stated direction is currently supported, \Reverse{} asks whether the opposite polarity is, and neither question can be answered by scaling a single signed exposure. The feasibility scales $\alpha_t^{\pm}$ are enforced before any probability is assigned, so the learning problem is posed over three implementable portfolios at equal requested magnitude, and the difference between the active pair is purely directional.

\subsection{Delayed cohorts and Safe-relative rewards}\label{sec:delayed_feedback}

None of the three candidates can be judged from a single next-day return, since their economic value is realized over a holding horizon of $H$ trading periods. A new cohort opens on each eligible date and is assigned cyclically to one of $H$ sleeves, so the structure is staggered rather than blockwise. Daily decision frequency is preserved while each directional decision is evaluated on its intended multi-period horizon, and at an interior date one cohort opens, one matures, and the remainder evolve unresolved; the implementation uses $H=5$. This structural delay differs from the generic delayed-feedback settings of \citet{WeinbergerOrdentlich2002Delayed} and \citet{JoulaniEtAl2013Delayed}, where delay is an exogenous nuisance attached to an otherwise immediate problem. Here the delay is the evaluation horizon itself, and the analysis later exploits its exact structure rather than paying a generic reduction. Maturity, reward, and normalization are
\begin{equation}
\label{eq:cohort_reward}
m_s=t_s+H-1,
\qquad
\ell_{s,a}=\log\frac{F_{s,a}}{F_{s,\Safe}},
\qquad
x_{s,a}=\frac{\ell_{s,a}}{B}\in[-1,1],
\qquad
x_{s,\Safe}=0,
\end{equation}
where $t_s$ is the opening date of cohort $s$, $F_{s,a}>0$ is the realized net wealth factor of candidate $a$ over the complete cohort $s$, trading costs included, and $B>0$ is the reward bound. Shadow accounts carry all three candidates to maturity even though only the authorized mixture trades, so the learner receives the full vector $x_s$ rather than the payoff of the executed mixture alone, and \Safe{} is the identical-zero numeraire of all learning and safety accounting. The bound $B$ is verified at every maturity instead of enforced by clipping, a property that later prices one unit of active probability at exactly one unit of normalized collateral.

\subsection{Share-adaptive signed master}\label{sec:signed_master}

Hard-coding one orientation would replace one unverified assumption with another, since a predictor useful in the \Follow{} direction during one period may later become uninformative or useful with the opposite polarity. The layer above the candidates therefore learns a probability vector over the three actions from matured rewards, under a Safe-biased prior that concentrates initial mass on the fallback and splits the remainder equally between the orientations. What this layer produces is a learning output, not an execution authorization, and that distinction carries the two safety layers that follow. The technical requirement is recoverability. After a long stretch in which one orientation earns nothing, probability must still be able to return to it once the polarity turns, which rules out plain exponential weighting. A Fixed-Share base learner \citep{HerbsterWarmuth1998Tracking} with share rate $\gamma$ applies the exponential reward update and then the share transition,
\begin{equation}
\label{eq:fixed_share}
\widetilde{q}^{(\gamma)}_{k,a}
=\frac{q^{(\gamma)}_{a}[k-1]\,e^{\eta x_{k,a}}}{\sum_{b}q^{(\gamma)}_{b}[k-1]\,e^{\eta x_{k,b}}},
\qquad
q^{(\gamma)}_{j}[k]=(1-\gamma)\,\widetilde{q}^{(\gamma)}_{k,j}+\frac{\gamma}{A},
\end{equation}
where $k$ counts matured rewards and $A=3$. The transition continually redistributes a fraction $\gamma$ of probability uniformly, so a deprecated action keeps a channel through which matured evidence can revive it. All bases share one action learning rate $\eta$ and differ only in $\gamma$; the architecture adapts the assumed switching intensity, not the step size.

Because a single $\gamma$ would encode a belief about how often the best action changes, the master runs $M$ bases on a share-rate grid and aggregates them from a uniform prior,
\begin{equation}
\label{eq:share_grid}
\begin{gathered}
\Gamma=\{0,\gamma_{\max}\}\cup\Bigl\{\tfrac{2^{j}}{T-1}:\ \tfrac{2^{j}}{T-1}<\gamma_{\max},\ j=0,1,\ldots\Bigr\},
\\
\bar{q}[k]=\sum_{m=1}^{M}v_m[k]\,q^{(\gamma_m)}[k],
\qquad
v_m[k]\propto v_m[k-1]\,e^{\beta y_{k,m}},
\end{gathered}
\end{equation}
where $y_{k,m}$ is the reward the pre-update distribution of base $m$ assigns to the matured vector and $\beta$ is the outer rate. The grid contains the static case, geometrically spaced responsive rates, and the pre-specified maximum, making switching intensity itself a learned quantity. The outer update is a Hedge aggregation \citep{FreundSchapire1997Hedge}, and it attains near-best-base performance without knowing the true switching intensity in advance; its regret price belongs to the theory. Writing $\bar{q}[k]$ for the aggregate state after $k$ matured rewards, with $\bar{q}[0]=\pi$, the raw probability supplied at the opening of cohort $s$ is $q_s^{\mathrm{raw}}=\bar{q}[\max\{0,s-H\}]$, since the cohorts matured before cohort $s$ opens are those numbered at most $s-H$. It is the master's answer to the second question, nothing more.

\subsection{Evidence gate and collateral budget}\label{sec:evidence_gate}

Between learning and execution stands a gate that observes only matured Safe-relative rewards. Fixed-Share intentionally preserves exploratory mass, so a positive raw probability need not mean the available evidence is strong. For each active action the gate computes an evidence summary and can transfer unsupported \Follow{} or \Reverse{} mass to \Safe{}; it never rejects \Safe{} itself and never creates active mass. This one-sidedness is a design invariant, recorded here because the analysis later charges safety interventions exactly the mass they remove. Rejection changes only the probability authorized for execution, and the learner underneath is not forced to forget, which preserves recovery after a genuine polarity change. The summary and the induced transformation are
\begin{equation}
\label{eq:gate}
\begin{gathered}
\mathcal{S}_{k,a}=\bigl(n_{k,a},\widehat{\mu}_{k,a},\underline{\mu}_{k,a}\bigr),
\qquad
q^{\mathrm{gate}}_{s,a}=\chi_{k,a}\,q^{\mathrm{raw}}_{s,a},
\quad a\in\{\Follow,\Reverse\},\quad k=\max\{0,s-H\},
\\
q^{\mathrm{gate}}_{s,\Safe}=1-q^{\mathrm{gate}}_{s,\Follow}-q^{\mathrm{gate}}_{s,\Reverse},
\end{gathered}
\end{equation}
where $n_{k,a}$ counts observations, $\widehat{\mu}_{k,a}$ is the empirical reward estimate, $\underline{\mu}_{k,a}$ is a one-sided lower-bound statistic, and $\chi_{k,a}\in\{0,1\}$ accepts action $a$ when the configured criterion on $\underline{\mu}_{k,a}$ passes, all computed from the $k$ rewards matured before cohort $s$ opens. The finite-sample calibration of the lower bound is part of the configuration reported with the experimental design. When neither orientation passes, the gated vector is the unit mass on \Safe{}, full abstention, which is distinct from reversing and requires no evidence at all.

The gate ignores losses still realizable by cohorts already open. Under overlapping holdings several active decisions coexist unresolved, so a sequence of individually admissible decisions can stack aggregate worst-case exposure that no single decision would have been permitted. Evidence answers whether an orientation is supported; it says nothing about whether the account can finance another unit of it. The collateral budget prices that remaining question,
\begin{equation}
\label{eq:collateral}
a(q)=q_{\Follow}+q_{\Reverse},
\qquad
q^{\top}x_s\geq-a(q),
\qquad
a_s^{\mathrm{exec}}=\min\bigl\{a_s^{\mathrm{gate}},\,b_s^{\mathrm{avail}}\bigr\},
\qquad
c_s=a_s^{\mathrm{exec}},
\end{equation}
with the active coordinates of $q_s^{\mathrm{gate}}$ scaled by the common factor $\zeta_s=a_s^{\mathrm{exec}}/a_s^{\mathrm{gate}}$, taken as $\zeta_s=1$ when $a_s^{\mathrm{gate}}=0$, and the removed mass transferred to \Safe{}. At maturity the settled reward of the cohort is $g_s=(q_s^{\mathrm{exec}})^{\top}x_s$, the quantity the reserve is credited with. Because the \Safe{} reward is identically zero and every active reward is bounded below by $-1$, one unit of active probability requires at most one unit of normalized collateral, which is why the normalization in \eqref{eq:cohort_reward} was chosen. The commitment $c_s$ is reserved at opening, before capital is exposed, and released only at maturity, when the realized reward of the stored executed probability is added to the reserve; the available budget $b_s^{\mathrm{avail}}$ therefore reflects delayed feedback instead of presuming unresolved outcomes. Common scaling preserves the learned composition between \Follow{} and \Reverse{}, so the budget changes how much directional risk is financed, never which direction is preferred. Gains above the operating cap are banked outside the spendable reserve, and the complete bookkeeping is recorded in Appendix~\ref{app:pseudocode}.

\subsection{Protected execution and the decision cycle}\label{sec:protected_execution}

Execution translates authorized probability into account structure. A fixed fraction $\rho$ of initial capital is tied to the \Safe{} strategy as a protected core, and the remainder funds the adaptive satellite that implements the executed mixtures; the implementation uses $\rho=0.80$. The fraction is not the master's Safe prior. The prior governs learning probabilities inside the satellite, whereas $\rho$ caps at account level how much capital adaptive decisions may touch at all, and confusing the two would collapse a capital allocation into a belief. The benchmark the core follows is itself staggered,
\begin{equation}
\label{eq:protected_core}
S_t=\sum_{h=1}^{H}S_{t,h},
\quad
S_{0,h}=\frac{1}{H},
\quad
C_t=\rho S_t,
\quad
V_t=\sum_{h=1}^{H}V_{t,h},
\quad
V_{0,h}=\frac{1-\rho}{H},
\quad
W_t^{\mathrm{prot}}=C_t+V_t,
\end{equation}
where sleeve $h$ of the induced \Safe{} benchmark $S_t$ implements the \Safe{} target on the same cohort timing and pays the same proportional costs as every other account, $V_t$ is the wealth of the satellite's $H$ sleeves, which trade the executed mixtures, and $W_t^{\mathrm{prot}}$ is protected wealth, the sum of the core and the satellite. The fallback is an implementable costed account, not a frictionless reference, and the protected core inherits its path scaled by $\rho$.

The daily cycle runs in a fixed order, stated as Algorithm~\ref{alg:controller} in Appendix~\ref{app:pseudocode}. The opening phase uses only pre-return information to construct the candidates, query the master, apply the gate and the budget, and open the cohort with its reserved commitment. The current return is then realized and every open sleeve is marked forward. Only after the new decision has been fixed are same-date maturities settled, in the order budget, master, gate, each consuming the probability vector stored at opening where settlement requires one. A cohort maturing on date $t$ therefore cannot influence the cohort opened earlier on date $t$. This causality claim is checked directly by the implementation audits of Section~\ref{sec:baselines}. The complete opening map composes the pipeline of Figure~\ref{fig:architecture} into one operator,
\begin{center}
\principle{candidates $\rightarrow$ $q_s^{\mathrm{raw}}\equiv\widehat q_s$ $\rightarrow$ $q_s^{\mathrm{gate}}$ $\rightarrow$ $q_s^{\mathrm{exec}}$ $\rightarrow$ $w_s^{\mathrm{exec}}=\sum_{a}q_{s,a}^{\mathrm{exec}}w_s^{a}$ $\rightarrow$ cohort opening.}
\end{center}
The map fixes what the architecture now owes. The executed probability path, the collateral reserve state, and the protected core are defined objects with exact bookkeeping, produced on every date by one learning decision and two one-sided interventions. The certificates these objects admit, tracking for the path and pathwise safety for the reserve and the core, are the subject of Section~\ref{sec:theory}.

\section{Theoretical Guarantees}\label{sec:theory}

The analysis is deterministic conditional on the realized bounded reward sequence. Nothing below assumes that the predictor carries positive expected alpha, that one orientation is permanently correct, or that returns follow a specified stochastic model. Throughout, $T$ counts economic reward rounds, $\mathcal{A}=\{\Safe,\Follow,\Reverse\}$ is the action set with $A=|\mathcal{A}|=3$, distinct from the number of tradable assets, $\Delta_A$ is the probability simplex over $\mathcal{A}$, and the results split into two branches. The tracking branch bounds the executed probability path against switching comparators, paying named penalties for delay, numerics, and safety intervention. The safety branch bounds cumulative Safe-relative loss pathwise, from collateral committed before unresolved outcomes are known. Both branches rest on six standing assumptions, stated once.

\begin{assumption}[Architecture invariants]\label{asm:architecture}
\mbox{}
\begin{itemize}
\item[(A1)] Matured rewards satisfy $x_s\in[-1,1]^{3}$ with $x_{s,\Safe}=0$, verified at every maturity.
\item[(A2)] All Fixed-Share bases share the learning rate $\eta>0$ and prior $\pi$, differ only in their share rates $\gamma_1,\ldots,\gamma_M$, and are aggregated by an outer Hedge update with rate $\beta>0$ from a uniform prior.
\item[(A3)] Write $\bar{q}_t:=\bar{q}[t-1]$ for the distribution the immediate-feedback master plays at round $t$, so $\bar{q}_1=\pi$. The decision distribution of cohort $s$ is the exact earlier master state $q_s^{\mathrm{dec}}=\bar{q}_{\kappa(s)}$ with $\kappa(s)=\max\{1,\,s-H+1\}$, which is $q_s^{\mathrm{raw}}$ of Section~\ref{sec:signed_master} in exact arithmetic, and no same-date maturity influences it.
\item[(A4)] Let $\widehat q_s\in\Delta_A$ denote the finite-precision value of $q_s^{\mathrm{dec}}$ that the implementation computes and hands to the gate, the implementation-facing pre-intervention probability. Gate and budget are one-sided\textup{:}
\[
0
\leq
q^{\mathrm{exec}}_{s,a}
\leq
q^{\mathrm{gate}}_{s,a}
\leq
\widehat q_{s,a},
\qquad
a\neq\Safe,
\]
with all removed active mass transferred to \Safe{}.
\item[(A5)] Every opened cohort commits $c_s\leq\mathsf{R}-\mathsf{P}$, the unencumbered reserve at opening; every settled reward $g_s=(q_s^{\mathrm{exec}})^{\top}x_s$ satisfies $g_s\geq-c_s$; replenishments $\xi_s\geq0$; withdrawals $\omega_s\geq0$ leave all outstanding commitments collateralized; and the initial reserve satisfies $\mathsf{R}_0\geq0$.
\item[(A6)] Executed cohort factors obey the exact split-account identity $F_s^{\mathrm{exec}}=\sum_a q^{\mathrm{exec}}_{s,a}F_{s,a}$ on the positive wealth domain, with $S_t>0$ and satellite wealth $V_t\geq0$.
\end{itemize}
\end{assumption}

Each item names the mechanism of Section~\ref{sec:methodology} that enforces it. (A1) is the checked reward normalization, (A2) the master construction, (A3) the two-clock cycle, (A4) the gate and budget transformations, (A5) the collateral bookkeeping, (A6) the split-account execution. The assumptions are therefore architecture properties to verify, not statistical hypotheses to believe, and the implementation audits them directly on the realized path.

\subsection{Share-adaptive tracking}\label{sec:theory_tracking}

The learning branch is stated first on the immediate-feedback clock, on which the matured vector $x_t$ is available before the next master distribution is formed, isolating learning from delay. The benchmark is an arbitrary comparator path $a_{1:T}\in\mathcal{A}^{T}$ that may switch among \Safe{}, \Follow{}, and \Reverse{} at any time, a strictly stronger standard than the best fixed action. Each Fixed-Share base prices such a path by the switching structure it exhibits,
\begin{equation}
\label{eq:fixed_share_path_complexity}
\mathcal{C}_{\gamma}(a_{1:T})
=-\log\pi_{a_1}-\sum_{t=2}^{T}\log P_{\gamma}\bigl(a_t\mid a_{t-1}\bigr),
\qquad
P_{\gamma}(j\mid i)=(1-\gamma)\1\{j=i\}+\frac{\gamma}{A}.
\end{equation}
The complexity $\mathcal{C}_{\gamma}$ is the negative log probability the Fixed-Share Markov prior with rate $\gamma$ assigns to the whole path, so persistent paths are cheap under small $\gamma$ and frequently switching paths are cheap under large $\gamma$. The economically right switching intensity is unknowable in advance, which is what the grid aggregation of \eqref{eq:share_grid} is for, and the minimum over the grid inside the bound below is its payoff.

\begin{theorem}[Share-adaptive tracking]\label{thm:share_adaptive_tracking}
Under \textup{(A1)}--\textup{(A2)}, for every bounded reward sequence and every comparator path $a_{1:T}\in\mathcal{A}^{T}$, the raw outer master $\bar{q}_t$ satisfies
\begin{equation}
\label{eq:share_adaptive_tracking_bound}
\sum_{t=1}^{T}\bar{q}_t^{\top}x_t
\;\geq\;
\sum_{t=1}^{T}x_{t,a_t}
-\min_{1\leq m\leq M}\frac{\mathcal{C}_{\gamma_m}(a_{1:T})}{\eta}
-\frac{\eta T}{2}
-\frac{\log M}{\beta}
-\frac{\beta T}{2}.
\end{equation}
\end{theorem}

The bound reads term by term. The comparator's reward is trailed by at most its best-grid path-complexity price over $\eta$, the learning-rate cost $\eta T/2$ of the base updates, and the two grid-overhead terms, which are the price of competing simultaneously with all $M$ switching models rather than selecting a share rate in advance. For a path with $K$ switches the geometric grid keeps the complexity price at $O(K\log(AT/K))$, and for a static path it collapses to $-\log\pi_{a_1}$, so the usual fixed-action benchmark is retained. The machinery is Fixed-Share tracking \citep{HerbsterWarmuth1998Tracking} under an outer Hedge aggregation \citep{FreundSchapire1997Hedge}; the proof, including the grid-complexity bound behind the explicit switch price, is in Appendix~\ref{app:proofs}. What the theorem does not say matters equally. It is a learning guarantee on Safe-relative rewards, not an alpha guarantee, and if every orientation performs poorly against \Safe{} it manufactures nothing.

\subsection{Delay and one-sided intervention}\label{sec:theory_delay}

The bound above governs an idealized clock, whereas the implementation decides while several cohorts remain unresolved. Under (A3) the distribution available at the opening of cohort $s$ is exactly an earlier state of the same master, not an approximation and not a delay-adjusted learner. For $H=1$ the gap vanishes. For $H>1$ the realized cost of staleness is the $\ell_1$ variation between the immediate and available states,
\begin{equation}
\label{eq:exact_delay_penalty}
q_s^{\mathrm{dec}}=\bar{q}_{\kappa(s)},
\qquad
\kappa(s)=\max\{1,\,s-H+1\},
\qquad
\Delta_T^{\mathrm{exact}}=\sum_{s=1}^{T}\bigl\|\bar{q}_s-\bar{q}_{\kappa(s)}\bigr\|_1.
\end{equation}

\begin{theorem}[Exact delayed-decision reduction]\label{thm:exact_delay}
Under \textup{(A1)} and \textup{(A3)}, for every bounded reward sequence,
\begin{equation}
\label{eq:exact_delay_reduction}
\sum_{s=1}^{T}\bigl(q_s^{\mathrm{dec}}\bigr)^{\top}x_s
\;\geq\;
\sum_{s=1}^{T}\bar{q}_s^{\top}x_s-\Delta_T^{\mathrm{exact}}.
\end{equation}
\end{theorem}

The reduction is an identity-grade statement, an additive and realized penalty rather than an inflation factor; the short proof is in Appendix~\ref{app:proofs}. Generic delayed-feedback reductions inflate regret multiplicatively in the delay, on the order of the $\sqrt{d+1}$ factor of \citet{WeinbergerOrdentlich2002Delayed} or the additive-per-round penalties of \citet{JoulaniEtAl2013Delayed}, because they must hold for arbitrary delay processes. The cohort structure is not arbitrary. Decisions consume exact earlier states of one master, so the entire cost of delay is the master's realized movement across the unresolved window, at most $(H-1)$ times its one-step variation, and $\Delta_T^{\mathrm{exact}}$ is measurable from the recorded probability path, as Section~\ref{sec:results_intervention} reports.

Even the delayed state is not what allocates capital, since the gate and the budget may still remove active mass. Their joint action is summarized by one structural lemma.

\begin{lemma}[One-sided gate-and-budget intervention]
\label{lem:one_sided_intervention}
Under \textup{(A1)} and \textup{(A4)}, define the active mass removed
from the implementation-facing probability by
\[
I_s
=
\sum_{a\neq\Safe}
\left(
\widehat q_{s,a}
-
q^{\mathrm{exec}}_{s,a}
\right)
\geq0,
\qquad
\mathcal I_T
=
\sum_{s=1}^{T}I_s.
\]
Then
\begin{equation}
\label{eq:one_sided_l1_identity}
\left\|
q_s^{\mathrm{exec}}
-
\widehat q_s
\right\|_1
=
2I_s,
\end{equation}
and
\begin{equation}
\label{eq:cumulative_one_sided_reward_bound}
\sum_{s=1}^{T}
\bigl(q_s^{\mathrm{exec}}\bigr)^{\top}x_s
\;\geq\;
\sum_{s=1}^{T}
\widehat q_s^{\top}x_s
-
\mathcal I_T.
\end{equation}
\end{lemma}

The content is the coordinate order of (A4). Safety intervention can cost at most the active mass it removes, because removed mass lands on the zero-reward \Safe{} coordinate and active rewards are bounded by one, and it can never manufacture exposure. Conservatism consequently enters every later bound as the single additive term $\mathcal{I}_T$, cumulative directional probability deliberately withheld; its proof is in Appendix~\ref{app:proofs}, and its economic reading waits until the certificate is assembled.

\subsection{Collateral safety}\label{sec:theory_collateral}

Tracking bounds cannot stop unresolved cohorts from stacking downside, since they compare against comparators rather than against capacity. The collateral account addresses this directly, reserving the full worst-case bounded loss of every cohort before it is funded instead of predicting losses probabilistically. Its events, openings and settlements indexed by $\tau$, need not coincide with reward rounds; $\mathsf{R}_\tau$ denotes the reserve, $\mathsf{P}_\tau$ the total commitment of open cohorts, $G_\tau$ the cumulative settled reward, $J_\tau$ the cumulative replenishment, and $\Omega_\tau$ the cumulative banked withdrawal, all in the normalized units of \eqref{eq:collateral}.

\begin{theorem}[Anytime delayed collateral safety]\label{thm:anytime_collateral}
Under \textup{(A1)} and \textup{(A5)}, solvency holds at every event, $\mathsf{R}_\tau\geq\mathsf{P}_\tau$, the reserve obeys the accounting identity $\mathsf{R}_\tau=\mathsf{R}_0+G_\tau+J_\tau-\Omega_\tau$, and consequently
\begin{equation}
\label{eq:anytime_completion_certificate}
B\bigl(G_\tau-\mathsf{P}_\tau\bigr)\;\geq\;-\mathsf{S}_\tau,
\qquad
\mathsf{S}_\tau:=B\bigl(\mathsf{R}_0+J_\tau\bigr).
\end{equation}
\end{theorem}

The certificate is pathwise and anytime. The term $-\mathsf{P}_\tau$ charges every unresolved cohort its full admissible loss, so the bound holds while feedback is still outstanding rather than only after the last settlement, and it holds under adversarial rewards because the only inputs are the solvency rules of (A5). That is a different notion of safety from the constrained and risk-sensitive control literature, where constraints hold in expectation or in distribution under a model of the environment \citep{GarciaFernandez2015SafeRL,AchiamEtAl2017CPO}. Here nothing is assumed about the reward process beyond boundedness, and the guarantee is an accounting invariant rather than a statistical one. The proof, an induction over opening and settlement events, is in Appendix~\ref{app:proofs}.

\begin{remark}[Protected-wealth floor]\label{rem:protected_floor}
Under \textup{(A6)}, $W_t^{\mathrm{prot}}=\rho S_t+V_t\geq\rho S_t$ on every admissible date $t$.
\end{remark}

The remark turns collateral units into an account-level wealth statement, and its two-line proof is in Appendix~\ref{app:proofs}. Protection holds datewise as well as at horizon, and it is the form Section~\ref{sec:results_protected_wealth} checks against the realized daily path.

\subsection{End-to-end guarantee}\label{sec:theory_end_to_end}

The composed result keeps two simultaneous lower bounds, a comparator-based tracking certificate and a comparator-free safety certificate, so realized performance is bounded by whichever is stronger on the realized path. This max-of-two-certificates form is the paper's thesis in mathematical shape. When predictions deserve trust the tracking arm is informative, and when they fail the safety arm binds without them.

\begin{theorem}[End-to-end safe share-adaptive tracking]\label{thm:end_to_end}
Let \textup{(A1)}--\textup{(A6)} hold, let $p_s=q_s^{\mathrm{exec}}$ denote the executed probability of cohort $s$, let $\widehat{q}_s$ be the finite-precision implemented vector of \textup{(A4)} with numerical discrepancy $\mathcal{E}_T^{\mathrm{num}}=\sum_{s=1}^{T}\|\widehat{q}_s-q_s^{\mathrm{dec}}\|_1$ from the exact state, and collect the tracking penalties as
\begin{equation*}
\Pi_T(a_{1:T})
=\min_{1\leq m\leq M}\frac{\mathcal{C}_{\gamma_m}(a_{1:T})}{\eta}
+\frac{\eta T}{2}
+\frac{\log M}{\beta}
+\frac{\beta T}{2}
+\Delta_T^{\mathrm{exact}}
+\mathcal{E}_T^{\mathrm{num}}
+\mathcal{I}_T.
\end{equation*}
Then, for every comparator path $a_{1:T}\in\mathcal{A}^{T}$, the realized Safe-relative cohort log rewards $L_s=\log\bigl(F_s^{\mathrm{exec}}/F_{s,\Safe}\bigr)$ satisfy
\begin{equation}
\label{eq:end_to_end_max_certificate}
\boxed{\;
\sum_{s=1}^{T}L_s
\;\geq\;
\max\Biggl\{\,-\mathsf{S}_T,\;\;
B\Bigl[\sum_{s=1}^{T}x_{s,a_s}-\Pi_T(a_{1:T})\Bigr]\Biggr\},
\;}
\end{equation}
with $\mathsf{S}_T=B(\mathsf{R}_0+J_T)$, where $G_T$ and $J_T$ are the cumulative settled reward and replenishment at the settlement event of cohort $T$, and independently $W_t^{\mathrm{prot}}\geq\rho S_t$ on every admissible date $t$.
\end{theorem}

The certificate is a consistency--robustness contract in the sense of algorithms with predictions \citep{LykourisVassilvitskii2018Caching,MitzenmacherVassilvitskii2020AlgPred}, transported to capital allocation. Consistency is the tracking arm, competitive with any switching use of the advice when the advice is good. Robustness is the safety arm, a hard bound paid for in advance that owes nothing to prediction quality. The contract language maps term by term onto the maximum, with one addition the caching setting does not need, a pathwise safety arm priced in reserved collateral and carrying a daily wealth floor under the split account.

\begin{proof}
Theorem~\ref{thm:share_adaptive_tracking} bounds the immediate-feedback master, $\sum_t\bar{q}_t^{\top}x_t\geq\sum_t x_{t,a_t}-\min_m\mathcal{C}_{\gamma_m}/\eta-\eta T/2-\log M/\beta-\beta T/2$, for any comparator path. Theorem~\ref{thm:exact_delay} replaces $\bar{q}_s$ by the causally available state $q_s^{\mathrm{dec}}$ at the additional cost $\Delta_T^{\mathrm{exact}}$. Because $\|x_s\|_{\infty}\leq1$, replacing $q_s^{\mathrm{dec}}$ by the implemented vector $\widehat{q}_s$ costs at most $\|\widehat{q}_s-q_s^{\mathrm{dec}}\|_1$ per cohort, hence $\mathcal{E}_T^{\mathrm{num}}$ in total, and this term vanishes in exact arithmetic. Lemma~\ref{lem:one_sided_intervention} transfers the bound from $\widehat{q}_s$ to the executed $p_s$ at cost $\mathcal{I}_T$, because the gate and budget only remove active mass and transfer it to \Safe{}. Chaining the four steps gives $\sum_s p_s^{\top}x_s\geq\sum_s x_{s,a_s}-\Pi_T(a_{1:T})$. The split-account identity of (A6) and concavity of the logarithm then give, for every cohort, $L_s=\log\sum_a p_{s,a}(F_{s,a}/F_{s,\Safe})\geq\sum_a p_{s,a}\log(F_{s,a}/F_{s,\Safe})=Bp_s^{\top}x_s$, which scales the linear bound into the tracking arm of \eqref{eq:end_to_end_max_certificate}. Independently, Theorem~\ref{thm:anytime_collateral} bounds the collateral surrogate at the settlement event of cohort $T$, where $G_T=\sum_{s\leq T}g_s$ because cohorts settle in opening order and $\mathsf{P}_\tau\geq0$ gives $BG_T\geq-\mathsf{S}_T$, and the same log-concavity step dominates the surrogate by the realized rewards, giving the safety arm. Both arms bound the same realized sum, so their maximum does. The floor is Remark~\ref{rem:protected_floor}.
\end{proof}

The proof is the architecture read in order, tracking through delay through numerics through intervention through solvency, each mechanism contributing exactly one named term. The intervention penalty deserves its own reading. $\mathcal{I}_T$ is exactly the active reward foregone when the safety layers refuse exposure, measured in withheld probability, so a large value need not indicate defective learning; it may record the deliberate price of abstention, which can be favorable when refused orientations would have lost, a possibility the abstention literature prices formally \citep{NeuZhivotovskiy2020Abstention}. Every penalty in $\Pi_T$ is measurable from recorded quantities on the realized path, and Section~\ref{sec:results_intervention} reports $\Delta_T^{\mathrm{exact}}$ and $\mathcal{I}_T$ from the recorded probability path.

\section{Experimental Design}\label{sec:experimental_design}

Every empirical claim below arrives pre-classified by a hierarchy of three evidence stages and an implementation-audit tier, fixed before results were interpreted. Development on a 24-stock universe built the architecture; a frozen cross-universe evaluation on 82 disjoint stocks tests whether the pre-specified directional rule transfers; post-holdout analysis diagnoses mechanisms. The hierarchy exists because the research question itself emerged from a negative transfer result, so development evidence can never double as confirmation of the controller that failure motivated. The severity is deliberate. Backtest overfitting makes it easy to report a high in-sample discovery whose out-of-sample expectation is negative \citep{BaileyEtAl2016PBO}, and a survey of 167 reinforcement-learning trading studies finds evaluation weaknesses of exactly this kind endemic \citep{PippasLudvigTurkay2025RLSurvey}. Table~\ref{tab:evidence_hierarchy} fixes each stage's universe, purpose, and evidential status.

\begin{table}
\caption{Empirical evidence hierarchy. `Frozen holdout' denotes an asset-universe holdout over the same calendar window as development.}
\label{tab:evidence_hierarchy}
\begin{center}
\begin{tabular}{p{0.17\textwidth}p{0.13\textwidth}p{0.36\textwidth}p{0.21\textwidth}}
\toprule
Stage & Universe & Primary purpose & Evidential status \\
\midrule
Development & 24 stocks & Construct and diagnose the prediction-assisted pipeline; compare baselines, predictor classes, and uncertainty-aware execution. & Development evidence; not independent confirmation. \\[1ex]
Frozen cross-universe holdout & 82 stocks & Test whether the pre-specified full-Ridge signal and its fixed \Reverse{} orientation transfer to a distinct asset universe. & Confirmatory test of the fixed transfer specification. \\[1ex]
Post-holdout signed analysis & 82 stocks & Diagnose polarity instability with signed orientation rules, abstention, and the complete adaptive controller. & Exploratory mechanism evidence; not confirmatory alpha evidence. \\[1ex]
Theory and implementation audits & Synthetic tests and realized paths & Verify algebraic identities, delayed chronology, one-sided intervention, solvency, and the protected-wealth invariant. & Implementation validation; not a return test. \\
\bottomrule
\end{tabular}
\end{center}
\end{table}

Two rules in the table bind everything downstream. Holdout means disjoint assets over the same calendar history, a test of cross-universe transfer rather than of performance in a later regime. And nothing examined after the frozen result was known can be promoted to confirmatory status, however well it performs. These are commitments, not caveats, and the results are reported against them.

\subsection{Universes and data}\label{sec:universes}

The development universe contains 24 liquid U.S. equities drawn mainly from energy, materials, mining, and utilities, with daily market and accounting data from February 2014 through September 2025. After feature and training history requirements, the out-of-sample development period runs from February 7, 2017 through September 18, 2025 and contains 2{,}166 sequential decisions, the same calendar window the holdout below occupies. This sample built the original prediction-assisted architecture, calibrated its walk-forward procedures, and compared its execution rules against Equal Weight. It is a development sample rather than an untouched test.

The frozen holdout contains 82 cross-sector equities disjoint from the development set, on a daily panel from February 20, 2014 through September 19, 2025 under the same data construction. After warm-up and estimation requirements, the frozen evaluation holds 2{,}166 daily observations beginning February 7, 2017, which under the $H=5$ overlapping-cohort construction yield 2{,}162 completed cohorts, called episodes in the archived result tables, with decisions through September 12, 2025 and final maturities that month. Every parameter and the signal orientation were fixed before any holdout return path was inspected, and poor holdout performance is never used to relabel the fixed-orientation specification afterwards. Episode arithmetic is stated here once and reused unchanged by the results.

Both universes share one ordered dictionary of 40 features, 17 price-derived variables spanning multiple return horizons, momentum, market beta, and downside volatility, and 23 technical and accounting variables including RSI, MACD, ADX, ATR, and lagged fundamental ratios. Accounting values are lagged so nothing enters a decision before it is economically observable, and the dictionary's identity across universes is enforced by audit, so a transfer failure cannot be blamed on changed inputs. Feature configuration is itself an experimental variable worth controlling, with documented first-order effects on DRL portfolio outcomes \citep{Ayari2026FeatureConfig}.

\subsection{Predictors, targets, and controller configuration}\label{sec:features_predictors}

The development prediction target aligns the statistical problem with benchmark-relative allocation. Writing $r^{\mathrm{EW}}_{t}$ for the Equal-Weight return, the predictor estimates for each asset $i$ a purged multi-horizon excess return,
\begin{equation}
\label{eq:development_target}
y_{t,i}=\sum_{h=0}^{9}\omega_h\bigl(r_{t+h,i}-r^{\mathrm{EW}}_{t+h}\bigr),
\qquad
\omega_h\propto0.9^{h},
\end{equation}
projected onto the complement of the trailing sixty-day factor exposures so the advice carries no market or style tilt, and standardized datewise. Only targets whose ten days have fully matured before a decision date enter training or calibration, behind a five-day embargo. The model supplies direction; feasibility, concentration limits, and costs stay with the portfolio layer, the same separation the candidate construction enforces.

The development predictor is an expanding walk-forward Ridge regression on the 40 features with asset fixed effects, retrained every 126 trading days from a minimum history of 504 days with observations weighted by a 756-day recency half-life, which yields eighteen retraining dates over the development window. At each retraining date the most recent 126 dates are withheld from fitting and used for split-conformal calibration of a directional uncertainty region at the 75\% level, so the execution rules below can price how far the predicted direction may be trusted. The realized mean validation information coefficient over the window is 0.148, a weak but nonzero signal, which is the regime the controller is built for.

One further predictor appears only in the supplement. A pooled feature-only Ridge uses the same 40 features with no asset or universe identity, its ridge strength is chosen on the 24-stock universe alone, and the fitted sequence is applied unchanged to the 82-stock universe as a leave-one-universe-out transfer. Only the saved summary of its controller run survives in the archive, and the supplement labels it accordingly.

The frozen transfer predictor is the legacy-target full-Ridge signal with the pre-specified \Reverse{} orientation as primary specification, and the clean-target full-Ridge \Reverse{} variant as secondary. Both were stored, with their target-validity masks and walk-forward caches, before any holdout inspection, and their primary-versus-secondary status was assigned then, not after. The frozen protocol also fixed the confirmatory statistic, a one-sided stationary-bootstrap test at the 5\% level of annualized net log growth against Equal Weight, with 3{,}000 replications and block length 13.

\begin{table}[t]
\caption{Controller configuration used in the post-holdout signed
analysis of the 82-stock universe. These values define the reported
exploratory controller and do not carry confirmatory status.}
\label{tab:controller_config}
\begin{center}
\footnotesize
\begin{tabular}{p{0.26\textwidth}p{0.22\textwidth}p{0.44\textwidth}}
\toprule
Parameter & Mechanism & Reported value \\
\midrule

Holding horizon $H$
& cohorts \eqref{eq:cohort_reward}
& 5 \\

Reward bound $B$
& normalization \eqref{eq:cohort_reward}
& 0.05 \\

Concentration cap $w_{\max}$
& feasible set \eqref{eq:feasible_set}
& 0.20 \\

Overlay magnitude $\delta$
& candidates \eqref{eq:follow_reverse}
& 0.10 \\

Direction $d_t$
& candidates \eqref{eq:follow_reverse}
& rank-based, long top 20\%, short bottom 20\% \\

Share-rate grid
& master \eqref{eq:share_grid}
& $M=13$, $\gamma_{\max}=0.50$ \\

Action learning rate $\eta$
& master \eqref{eq:fixed_share}
& 0.031879 \\

Outer rate $\beta$
& master \eqref{eq:share_grid}
& 0.048711 \\

Safe prior $\pi_{\Safe}$
& master initialization
& 0.50 \\

Gate evidence rule
& evidence gate \eqref{eq:gate}
& last 50 matured rewards, at least 25, one-sided normal test at 5\% \\

Collateral reserve, replenishment, cap
& collateral budget \eqref{eq:collateral}
& 1, 1, 2 normalized units (0.05, 0.05, 0.10 in log wealth at $B=0.05$) \\

Protected fraction $\rho$
& split account \eqref{eq:protected_core}
& 0.80 \\

Cost coefficient $c$
& transaction-cost accounting
& 5 bp per unit turnover \\

\bottomrule
\end{tabular}
\end{center}
\end{table}

Table~\ref{tab:controller_config} consolidates the principal
controller-level configuration used for the reported signed analysis.
The learning rates follow closed-form rules in the expected number of
matured rewards $T=2{,}162$, the action count $A=3$, and the base
count $M$,
\begin{equation}
\label{eq:learning_rates}
\eta=\min\Bigl\{\tfrac12,\ \sqrt{2\log A/T}\Bigr\},
\qquad
\beta=\min\Bigl\{\tfrac12,\ \sqrt{2\log M/T}\Bigr\},
\end{equation}
which give the tabulated values, and the share-rate grid of
\eqref{eq:share_grid} contains the static base $\gamma=0$ and
terminates at $\gamma_{\max}=0.50$, giving $M=13$ bases. The signed
master is initialized with Safe prior $\pi_{\Safe}=0.50$, with the
remaining prior mass split across Follow and Reverse.

The evidence gate uses only fully matured bounded economic rewards.
For each active action it keeps the last 50 matured rewards, and once
at least 25 are available it computes their mean and a one-sided normal
lower bound at the 5\% level, the mean less 1.645 standard errors. The
action is retained when that bound is positive, and otherwise its
probability mass is transferred to Safe. The level is nominal, because
overlapping cohorts make consecutive rewards dependent and the standard
error does not correct for that, so the gate is a calibrated screen and
not a sequential guarantee, which is why the safety certificates never
rely on it. No current or partially matured cohort enters the decision.
The collateral account opens with a reserve of $\mathsf{R}_0=1$
normalized unit, receives a further $J_T=1$ unit of scheduled
replenishment spread evenly over the expected matured updates, and
banks gains above a cap of 2 units; at $B=0.05$ these are 0.05, 0.05,
and 0.10 in log wealth, so the full-horizon certified floor of
Theorem~\ref{thm:anytime_collateral} is $-\mathsf{S}_T=-B(\mathsf{R}_0+J_T)=-0.10$
in log units. The protected-execution layer operates according to the causal
sequence defined in Section~\ref{sec:protected_execution}.

Because the signed architecture was examined after the frozen transfer
result was known, these quantities characterize the post-holdout
exploratory mechanism analysis and are not presented as pre-holdout
confirmatory choices.

\subsection{Baselines, ablations, and audits}\label{sec:baselines}

The external benchmark in both universes is Daily Equal Weight, hereafter Equal Weight, long-only and fully invested, and every turnover-bearing strategy is charged the same drift-relative proportional cost, which is what makes the comparisons fair rather than an afterthought.

The development stage compares three execution rules that differ only in how much predictive advice they accept. Safe OPMD ignores advice entirely. Always Advice adds the predicted direction to the Safe gradient at a fixed strength of ten and executes the result. Certified execution evaluates, for each of five advice strengths, the convex mixture of the Safe target and the advised target whose split-conformal directional lower confidence bound is largest, with the Safe target itself always admissible so the selected bound can never be negative, and executes that mixture. The axis these rules trace, the price of accepting weak advice under conservative execution, is a development diagnostic, and certified execution is the ancestor of the evidence gate, not a test of the later signed controller.

The frozen 82-stock comparison holds Daily Equal Weight, Safe OPMD run as a single portfolio, the Matched Fixed Core, and the primary and secondary Ridge \Reverse{} specifications. The Matched Fixed Core holds three quarters of capital at the Safe target and one quarter at Equal Weight with no active overlay, under the same single-portfolio accounting, and each frozen \Reverse{} path is that core carrying a 0.10 $\ell_1$ overlay in the \Reverse{} orientation, so the core isolates what the overlay adds. The corresponding raw orientations are kept in the diagnostic record so an unfavorable \Reverse{} result cannot be read without its mirror. After the frozen verdict, the post-holdout analysis adds signed variants, fixed \Follow{} and \Reverse{} references, local and cumulative signed orientation, and consensus-abstention rules, and then assembles the complete architecture mechanism by mechanism, master, then gate, then budget, then split account, in the same order the pipeline composes them. Every one of these paths is exploratory by the hierarchy's second rule.

The implementation is audited directly, candidate feasibility and simplex preservation, realized overlay size, gate transfer identities, common collateral scaling, reward boundedness, delayed chronology, split-account aggregation, and the daily protected-wealth invariant. The audits verify that the code path that ran is the controller Section~\ref{sec:theory} analyzed, item-level results in the supplement; they are implementation checks, not tests of investment alpha.

Two further exploratory exercises reproduce the signed controller's reference path, the clean-target full-Ridge signal under the complete controller, with the mechanism altered. The first removes the gate, the budget, or both, and runs the three fixed actions \Safe{}, \Follow{}, and \Reverse{} through the same split account, so each layer's contribution can be read against the others. The second replaces the market with a twelve-asset generator in which a planted signal drives returns with a chosen strength through \Follow{}, neutral, and \Reverse{} regimes of equal length, at strengths 0, 0.03, 0.06, and 0.12 over ten seeds of 900 episodes each. Strength is a simulation input, not a market estimate, and nothing in either exercise was tuned.

\subsection{Costs, metrics, and inference}\label{sec:evaluation}

The baseline cost specification charges five basis points per unit of drift-relative turnover, the convention of \eqref{eq:feasible_set} with the coefficient recorded in Table~\ref{tab:controller_config}. The model is deliberately linear and transparent. Nonlinear market impact, bid--ask variation, order-book depth, and liquidity shocks are omitted, and the omissions are declared limitations rather than absorbed into an estimated parameter. Sensitivity of realized paths to the coefficient is examined in the results; only the convention lives here. Performance is measured from realized wealth paths,
\begin{equation}
\label{eq:performance_metrics}
\operatorname{CAGR}=\Bigl(\frac{W_n}{W_0}\Bigr)^{252/n}-1,
\qquad
\operatorname{MDD}=\min_{1\leq t\leq n}\Bigl\{\frac{W_t}{\max_{0\leq u\leq t}W_u}-1\Bigr\},
\qquad
\overline{\operatorname{TO}}=\frac{1}{n}\sum_{t=1}^{n}\operatorname{TO}_t.
\end{equation}
The measure set is completed by annualized volatility, the Sharpe ratio from daily net returns, final wealth, and, for the 82-stock analysis, terminal wealth and annualized log growth relative to Daily Equal Weight. Wealth-based CAGR is preferred over the annualized mean daily return because compounding, not arithmetic optics, is what the investor receives.

Two \Safe{} series appear. The benchmark tables use Safe OPMD run as one portfolio, whereas the controller exhibits measure against the induced \Safe{} benchmark of \eqref{eq:protected_core}, the same target carried in five staggered sleeves, whose CAGR differs from the single portfolio's by one basis point through sleeve accounting. The two are labeled wherever they appear and never substituted for each other.

Inference uses a paired stationary block bootstrap \citep{PolitisRomano1994} on daily net returns, 3{,}000 replications with block length $\lfloor n^{1/3}\rfloor=13$, so short-range serial dependence survives resampling, applied to Sharpe-ratio and log-growth differences. It is used exactly twice. In development it compares certified execution with Equal Weight, and in the frozen test it is the pre-registered confirmatory statistic of Section~\ref{sec:features_predictors}.

Three restrictions guard the interpretation. Non-rejection of a Sharpe difference is not equivalence, only a failure to detect. The development test lends nothing to the 82-stock architecture. And the confirmatory question of the frozen test is directional and economic, whether the pre-specified transfer yields a positive implementable edge, which the pre-registered test answers once and no later analysis can reopen. The results now answer the hierarchy's three questions in order, what development taught, whether the frozen specification transferred, and what the signed mechanisms did.

\section{Empirical Results}\label{sec:results}

The ordering below is intentional and follows the hierarchy of Table~\ref{tab:evidence_hierarchy}. The development evidence explains why a prediction-assisted overlay appeared worth studying. The frozen holdout then decides whether the cross-universe claim survives, and its verdict is not altered by anything that follows. The polarity analysis diagnoses the nature of the failure, and the mechanism results answer it. Nothing exploratory is staged as a comeback, by the second rule of Section~\ref{sec:experimental_design}.

\subsection{Development-universe baseline comparison}\label{sec:results_development}

Development provides no evidence that prediction assistance improves raw performance. In Table~\ref{tab:development_results}, Daily Equal Weight's 16.86\% CAGR exceeds Safe OPMD's 16.71\%, their Sharpe ratios sit at 0.775 against 0.779, and their maximum drawdowns are nearly identical. Certified execution coincides with Safe OPMD on every day of the sample, because the directional lower confidence bound never turned positive and the rule therefore never accepted advice, so its row is the Safe row. Leading with this non-result fixes the register for the whole section. The development experiment motivated an implementation idea, and it never established a return edge.

\begin{table}
\caption{Out-of-sample performance in the 24-stock development universe, net of the 5-bp baseline cost, at the precision of the archived output. Certified execution equals Safe OPMD because its certificate never accepted advice. This sample was used during method development and is not independent confirmation of the final signed controller.}
\label{tab:development_results}
\begin{center}
\footnotesize
\begin{tabular}{lrrrrrr}
\toprule
Strategy & CAGR & Ann.\ vol. & Sharpe & Max.\ DD & Avg.\ turn. & Final wealth \\
\midrule
Daily Equal Weight & 16.86\% & 23.79\% & 0.775 & $-44.53\%$ & 1.383\% & 3.8165 \\
Safe OPMD & 16.71\% & 23.37\% & 0.779 & $-44.49\%$ & 0.104\% & 3.7728 \\
Always Advice & 16.58\% & 23.35\% & 0.774 & $-44.42\%$ & 0.117\% & 3.7372 \\
Certified execution & 16.71\% & 23.37\% & 0.779 & $-44.49\%$ & 0.104\% & 3.7728 \\
\bottomrule
\end{tabular}
\end{center}
\end{table}

What the sample does show is economically distinctive. The conservative rules hold Equal Weight's return and drawdown profile while trading an order of magnitude less, 0.104\% average daily turnover against 1.383\%, a reduction of roughly 92\%. Always Advice trades more than the anchor at 0.117\% and earns less, so executing the prediction without a certificate buys nothing here. The benefit is controlled implementation, the ability to keep a prediction layer available without letting it force trading, and it is that property, not alpha, which later transfers.

One bootstrap test closes the development record under the restrictions of Section~\ref{sec:evaluation}. Certified execution against Equal Weight shows a Sharpe difference of $+0.004$ with a 95\% interval from $-0.048$ to $+0.055$ and a two-sided $p$-value of 0.93, and an annualized log-growth gap of $-0.13\%$ with an interval from $-1.36\%$ to $+1.07\%$. Non-rejection is not equivalence, and this test lends nothing to the 82-stock architecture.

\subsection{Frozen cross-universe evaluation}\label{sec:results_holdout}

What is frozen deserves restating before its verdict. The primary path is the legacy-target full-Ridge \Reverse{} specification, the secondary is the clean-target variant, and both directions, targets, and parameters were fixed before any 82-stock return path was seen, per Section~\ref{sec:features_predictors}.

\begin{table}
\caption{Frozen cross-universe results in the 82-stock universe at the 5-bp baseline cost, at the precision of the archived output. The final column is annualized realized log growth relative to Daily Equal Weight. `Reverse' corresponds to `Inverse' in the archived implementation.}
\label{tab:frozen_holdout_results}
\begin{center}
\footnotesize
\setlength{\tabcolsep}{4.5pt}
\begin{tabular}{lrrrrrr}
\toprule
Strategy & CAGR & Ann.\ vol. & Sharpe & Max.\ DD & Avg.\ turn. & Rel.\ EW \\
\midrule
Daily Equal Weight & 14.38\% & 18.67\% & 0.813 & $-38.58\%$ & 1.070\% & 0.00\% \\
Safe OPMD & 13.95\% & 18.16\% & 0.811 & $-37.28\%$ & 0.077\% & $-0.37\%$ \\
Matched Fixed Core & 14.06\% & 18.28\% & 0.812 & $-37.60\%$ & 0.282\% & $-0.27\%$ \\
Legacy Full-Ridge Reverse & 13.88\% & 18.28\% & 0.803 & $-38.03\%$ & 0.965\% & $-0.43\%$ \\
Clean Full-Ridge Reverse & 13.74\% & 18.23\% & 0.798 & $-37.92\%$ & 1.010\% & $-0.55\%$ \\
\bottomrule
\end{tabular}
\end{center}
\end{table}

The verdict column of Table~\ref{tab:frozen_holdout_results} is the last one. Daily Equal Weight finishes highest, with 14.38\% CAGR, terminal wealth 3.1723, and Sharpe 0.813, while the primary frozen \Reverse{} path loses 0.43\% annually against it in realized log growth, reaching roughly 96.4\% of the Equal-Weight terminal wealth, and the secondary specification is weaker still at $-0.55\%$. The pre-registered one-sided test returns $p=0.83$, with a 95\% interval for the annual log-growth gap from $-1.33\%$ to $+0.44\%$, so the confirmatory hypothesis is not supported. The conservative paths behave differently. Safe OPMD holds Sharpe 0.811 with turnover cut from 1.070\% to 0.077\% and a smaller drawdown, and the Matched Fixed Core sits similarly close, yet the two remain 37 and 27 basis points per year below Equal Weight in compounding. Closeness at far lower turnover transfers; a return edge does not.

This is the main negative empirical result of the paper. Low-turnover conservatism carries across universes, and the directional claim does not. The frozen evidence is never described here or later as confirming predictive alpha, and everything after this subsection is diagnosis and mechanism, reported under the exploratory label the hierarchy assigns it.

\subsection{Failure of a fixed signal orientation}\label{sec:results_polarity}

The failed transfer poses a diagnostic question with a tempting answer. Was the predictor economically useless, or merely mis-signed, in which case permanently flipping \Reverse{} to \Follow{} would repair it? The flip is examined directly and found insufficient.

\begin{table}
\caption{Fixed-orientation diagnostics in the 82-stock universe. These comparisons diagnose polarity and do not alter the outcome of the frozen primary test.}
\label{tab:fixed_orientation_results}
\begin{center}
\footnotesize
\setlength{\tabcolsep}{4.5pt}
\begin{tabular}{lrrrrr}
\toprule
Specification & CAGR & Sharpe & Max.\ DD & Avg.\ turn. & Rel.\ EW \\
\midrule
Legacy Follow & 14.04\% & 0.810 & $-37.21\%$ & 0.989\% & $-0.30\%$ \\
Legacy Reverse & 13.88\% & 0.803 & $-38.03\%$ & 0.965\% & $-0.43\%$ \\
Clean Follow & 14.16\% & 0.814 & $-37.32\%$ & 1.057\% & $-0.19\%$ \\
Clean Reverse & 13.74\% & 0.798 & $-37.92\%$ & 1.010\% & $-0.55\%$ \\
\bottomrule
\end{tabular}
\end{center}
\end{table}

Table~\ref{tab:fixed_orientation_results} shows \Follow{} beating \Reverse{} under both stored targets, which explains why the frozen \Reverse{} orientation failed, while every fixed orientation stays below Equal Weight, with Clean Follow still at $-0.19\%$ annualized relative log growth. Reversing the sign after observing the holdout would improve the point estimate without converting the experiment into evidence of positive transfer, and choosing the better sign from the same holdout would be exactly the post-hoc promotion the second rule of Section~\ref{sec:experimental_design} forbids.

\begin{figure}
\centering
\includegraphics[width=0.72\textwidth]{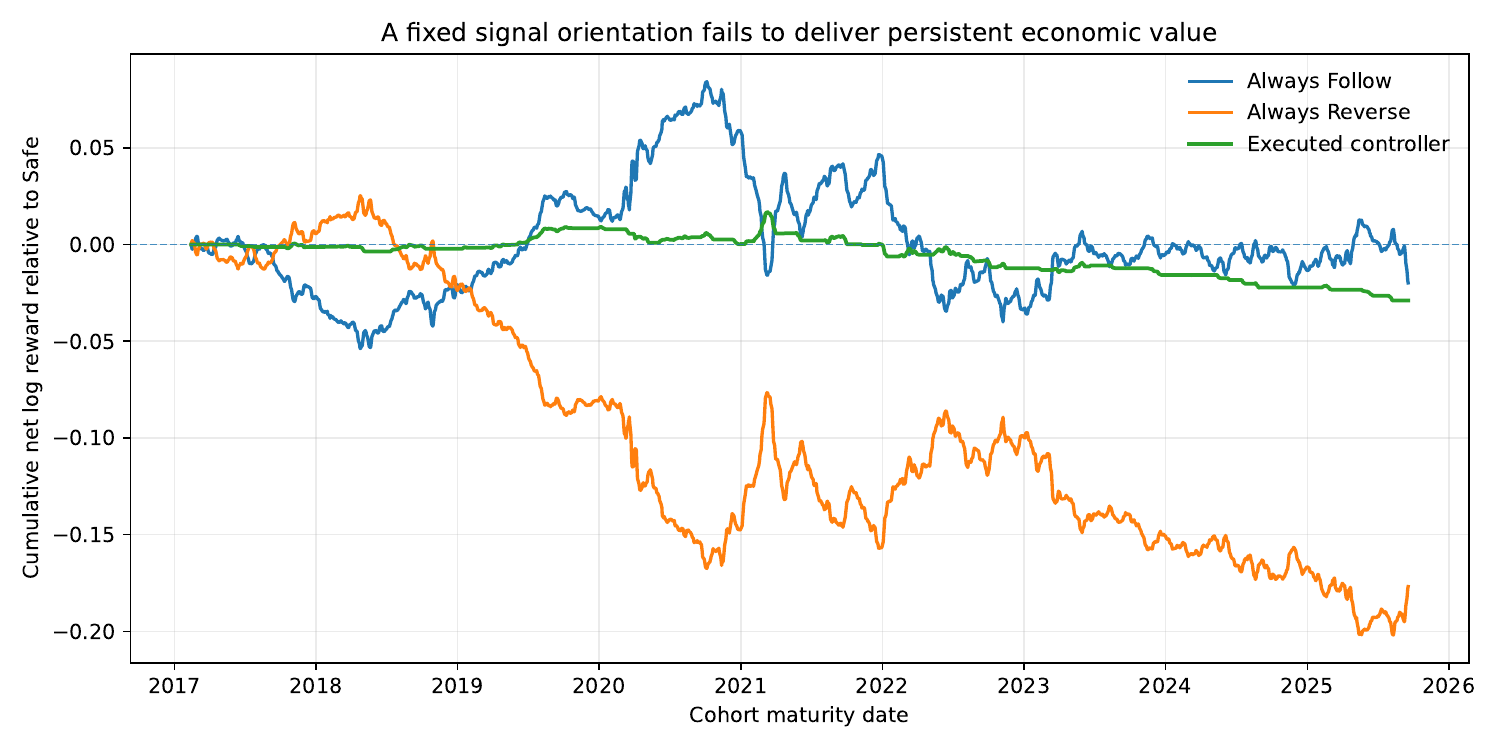}
\caption{A fixed signal orientation fails to deliver persistent economic value. Cumulative realized net log reward relative to \Safe{} at cohort maturities for Always Follow, Always Reverse, and the executed adaptive controller. The executed-controller path is a post-holdout mechanism diagnostic, not a confirmatory alpha result, and cohort-level cumulative reward is distinct from the daily protected-wealth ratio analyzed later.}
\label{fig:fixed_orientation_failure}
\end{figure}

Figure~\ref{fig:fixed_orientation_failure} adds what the full-sample table cannot show, and it teaches two lessons. Orientation value is time-varying, since Always Follow accumulates useful reward over parts of the sample and then surrenders it, so a sign correct over one interval need not stay correct over the next. And \Reverse{} is not hidden symmetric alpha, being substantially more damaging than \Follow{} on this path. There is also no tension between the figure and the table. The table reports daily portfolio wealth from complete trading paths, whereas the figure accumulates overlapping $H=5$ Safe-relative cohort rewards, a diagnostic learning quantity, not a second wealth measure. The finding compresses to one line,
\begin{equation}
\label{eq:prediction_polarity_failure}
\principle{Prediction can contain useful information \ $\not\Rightarrow$ \ its economic direction is stable.}
\end{equation}

The implication demands three capabilities no fixed-sign portfolio has. Allocation between \Follow{} and \Reverse{} must be learned from matured economic evidence, abstention must return probability to \Safe{} when neither orientation is supported, and funded exposure must be limited while unresolved decisions remain outstanding. The remaining results examine whether the controller delivers them, under a question that is narrower and more defensible than alpha. When prediction direction is uncertain, does the controller recognize that uncertainty and prevent a failed orientation from becoming an uncontrolled exposure?

\subsection{Safety intervention and abstention}\label{sec:results_intervention}

Even a share-adaptive master can propose active mass on weak evidence, since Fixed-Share deliberately preserves exploratory probability. The signed controller reported here and in the remaining real-data subsections runs on the clean-target full-Ridge signal, the predictor of the secondary frozen specification, with the legacy-target signal and the feature component retained as diagnostics in the supplement. The realized action of the two safety layers is therefore examined directly, with active probability $a(q_s)=q_{s,\Follow}+q_{s,\Reverse}$ as in \eqref{eq:collateral} so that reductions correspond exactly to mass transferred toward \Safe{}.

\begin{figure}
\centering
\includegraphics[width=0.72\textwidth]{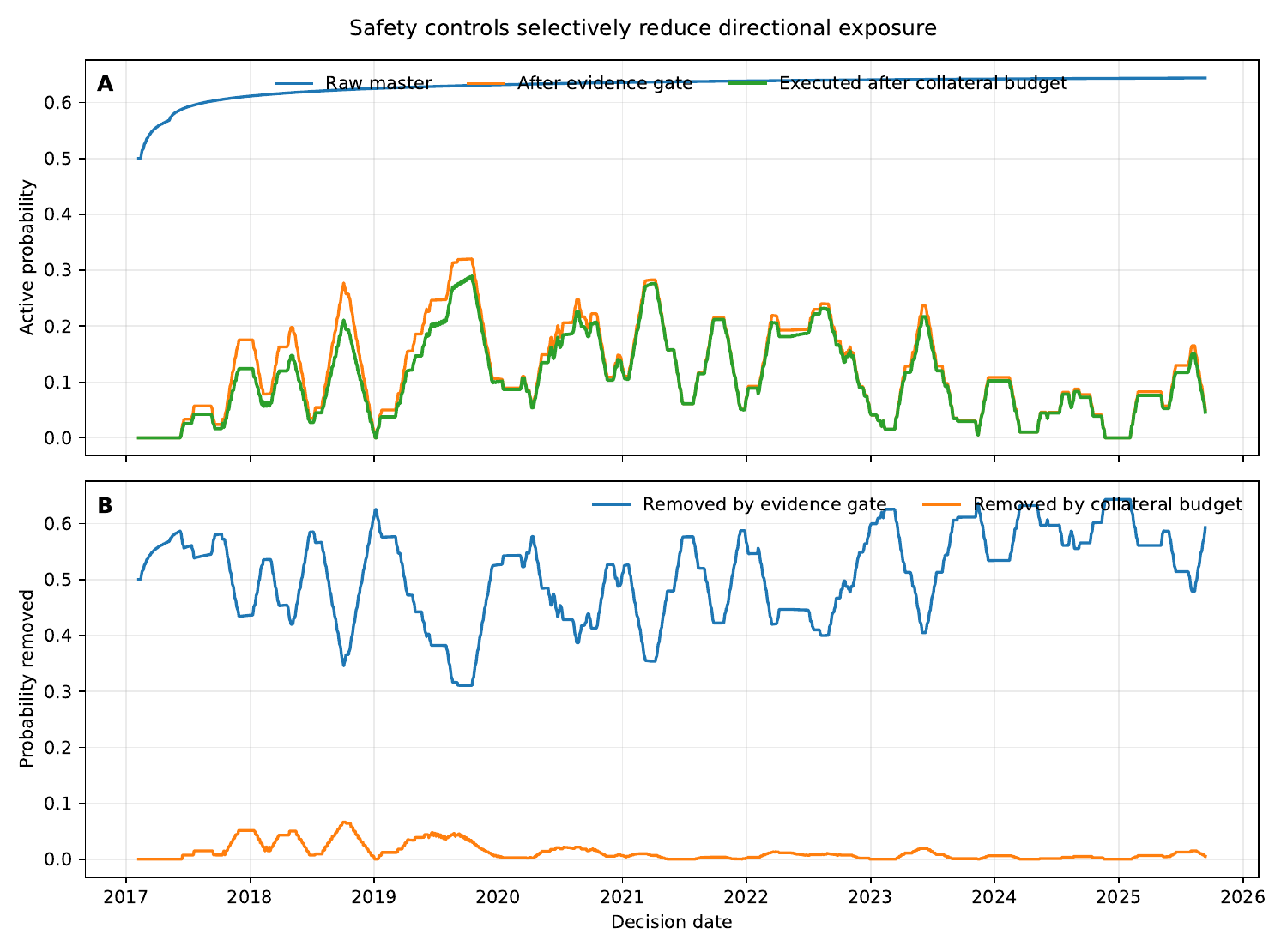}
\caption{Safety controls selectively reduce directional exposure. Panel A tracks active probability proposed by the raw master, remaining after the evidence gate, and executed after collateral authorization, as trailing 63-decision rolling means, since per-cohort values alternate as individual sleeves pass the gate. Panel B decomposes the removals by layer on the same smoothing. Removed probability is transferred to \Safe{}, and neither intervention can increase exposure. The signed controller is a post-holdout mechanism analysis, not a confirmatory alpha result.}
\label{fig:safety_intervention}
\end{figure}

Panel A of Figure~\ref{fig:safety_intervention} shows the raw master keeping most active mass, starting near one half and rising toward two thirds, exactly the recoverability Fixed-Share is built for. The executed path looks entirely different. Gating leaves directional exposure generally below one third and periodically at or near zero, stretches of full abstention in which the controller declines to force a choice between \Follow{} and \Reverse{} and reassigns the rejected mass to \Safe{}.

Panel B attributes the reduction. The gate dominates on this realized path, while the budget removes a generally smaller amount that binds when fundable and statistically admissible exposure differ, which is its designed role rather than redundancy. The implementation audits confirm the structure the theory requires, with no probe or realized trajectory ever increasing active probability and all transfer and scaling identities holding exactly; item-level audit records are in the supplement. The penalties of Theorem~\ref{thm:end_to_end} are read off the same recorded path. The intervention penalty is $\mathcal{I}_T=1{,}131$ normalized units over 2{,}162 cohorts, a mean of 0.52 of active probability withheld per decision, so $\mathcal{I}_T$ measures exposure the implemented system actually refused to fund, not an abstract regularizer. The exact delay penalty is $\Delta_T^{\mathrm{exact}}=2.74$, a mean $\ell_1$ movement of 0.0013 per cohort between the state the master would have played with immediate feedback and the state it did play, so staleness costs little on this path. The numerics term is at machine precision, the split-account identity holding to $10^{-16}$.

\subsection{Protected wealth relative to Safe}\label{sec:results_protected_wealth}

Reduced directional probability must still show up at account level. Protected wealth is $W_t^{\mathrm{prot}}=\rho S_t+V_t$ with $\rho=0.80$, measured against the induced five-sleeve \Safe{} benchmark $S_t$ of \eqref{eq:protected_core} rather than the single-portfolio Safe OPMD of Table~\ref{tab:frozen_holdout_results}, and Remark~\ref{rem:protected_floor} guarantees $W_t^{\mathrm{prot}}/S_t\geq0.80$ on every admissible date, a hard bound that permits a 2{,}000-basis-point relative shortfall. The floor is not a forecast, and the realized distance from it is the quantity of interest.

\begin{figure}
\centering
\includegraphics[width=0.72\textwidth]{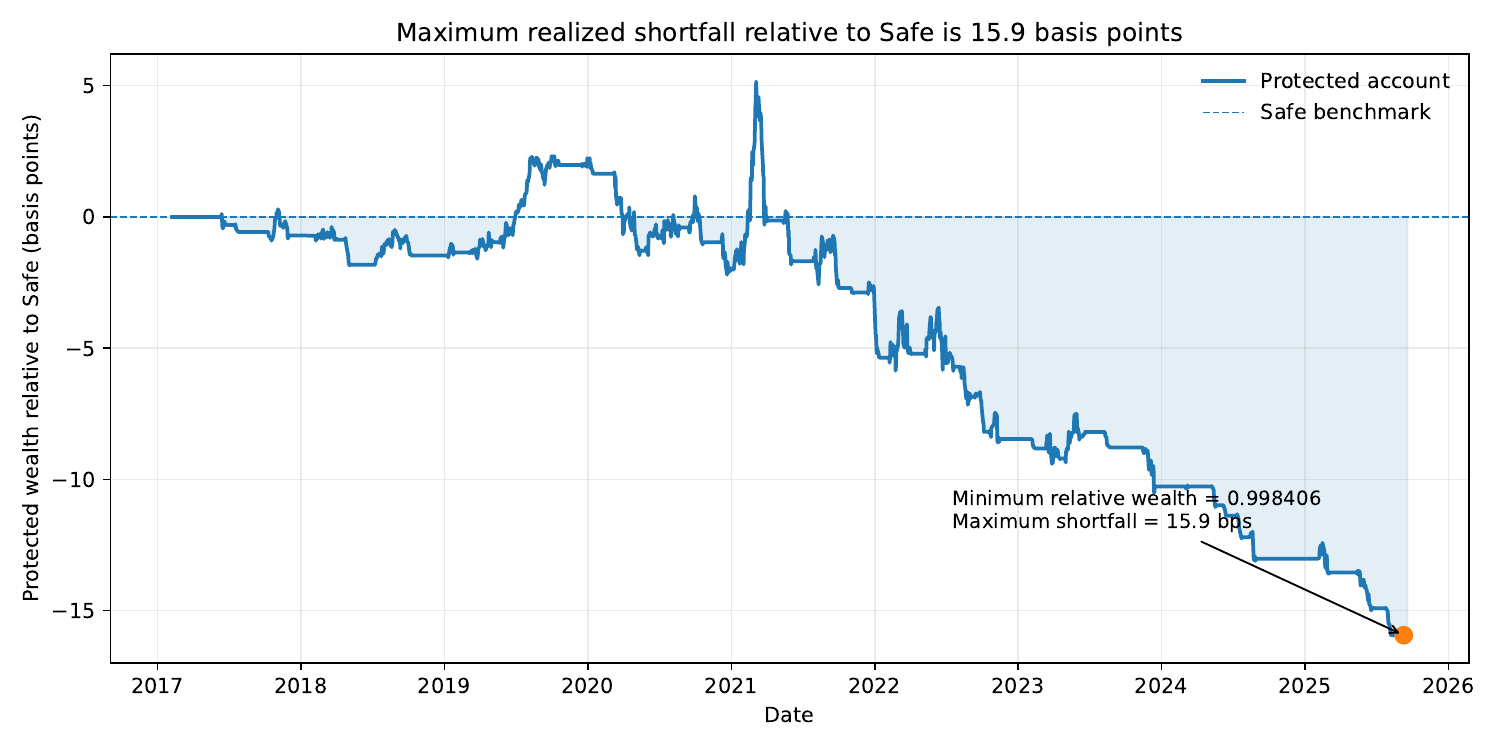}
\caption{Maximum realized shortfall relative to \Safe{} is 15.9 basis points. Daily protected split-account wealth relative to the induced \Safe{} benchmark, in basis points against zero. This daily wealth quantity is distinct from the cumulative overlapping cohort reward of Figure~\ref{fig:fixed_orientation_failure}.}
\label{fig:protected_wealth}
\end{figure}

Figure~\ref{fig:protected_wealth} shows excursions on both sides of \Safe{}, with the largest gap late in the sample. The realized extremum is
\begin{equation}
\label{eq:empirical_min_protected_ratio}
\min_t\frac{W_t^{\mathrm{prot}}}{S_t}=0.998406,
\qquad
10^{4}\bigl(1-0.998406\bigr)\approx15.9\ \text{basis points},
\end{equation}
against the certified worst case of 2{,}000. The gap between floor and realization does not discredit the theorem, which must hold under any prediction quality. The realized path stacks four conservatisms the worst case cannot assume, prior mass retained on \Safe{} by the master, unsupported mass removed by the gate, active exposure scaled by the collateral budget, and only $1-\rho=20\%$ of capital assigned to the satellite at all. Each layer reduced executed active mass in Figure~\ref{fig:safety_intervention}; what each contributed to the shortfall is measured in Section~\ref{sec:results_ablation}.

The claim this figure supports is bounded. \Safe{} itself dips to a wealth level of roughly 0.942 on the audited path, so the guarantee limits incremental damage from adding untrusted predictions to the chosen fallback, never market losses the fallback already bears. Within that reading it is the economically strongest safety result in the paper. After allowing the system to learn from \Follow{} and \Reverse{} outcomes throughout, realized wealth stayed within 15.9 basis points of the fallback path.

\subsection{Transaction-cost sensitivity}\label{sec:results_costs}

Trading intensity is the remaining implementation margin, and it is priced by a fixed-path repricing exercise. Decisions and turnover are frozen at their realized values and only the proportional cost coefficient varies, with no retraining, no orientation change, and no retuning at any cost level. The three repriced paths and their baseline characteristics are fixed in Table~\ref{tab:cost_sensitivity_paths}.

\begin{table}
\caption{Baseline paths used in the 82-stock transaction-cost sensitivity analysis. Consensus Abstention is a post-holdout exploratory strategy.}
\label{tab:cost_sensitivity_paths}
\begin{center}
\small
\begin{tabular}{lrr}
\toprule
Strategy & CAGR at 5 bp & Avg.\ turnover \\
\midrule
Daily Equal Weight & 14.375\% & 1.070\% \\
Frozen Reverse & 13.884\% & 0.965\% \\
Consensus Abstention (exploratory) & 14.168\% & 0.793\% \\
\bottomrule
\end{tabular}
\end{center}
\end{table}

\begin{figure}
\centering
\includegraphics[width=0.72\textwidth]{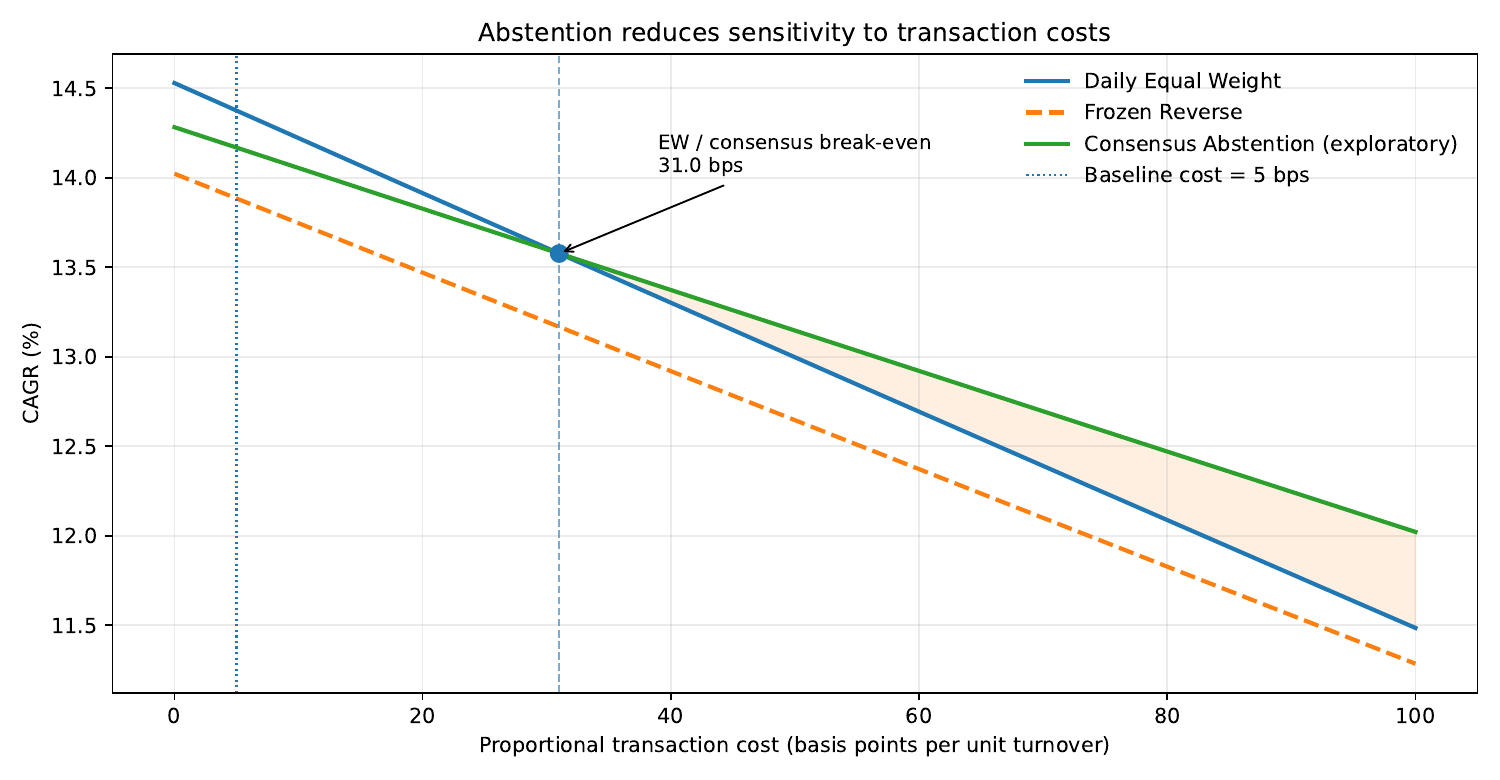}
\caption{Transaction-cost sensitivity under fixed realized portfolio paths, repriced from 0 to 100 basis points per unit turnover. The exploratory Consensus Abstention path reaches parity with Equal Weight near 31 basis points, reflecting its lower turnover rather than newly generated forecasting alpha. Frozen Reverse remains inferior throughout.}
\label{fig:transaction_cost_sensitivity}
\end{figure}

Figure~\ref{fig:transaction_cost_sensitivity} reprices the three paths. Equal Weight deteriorates fastest because it trades most, and the abstention path crosses it near 31 basis points per unit turnover. The reading is deliberately narrow. Gross decisions never change across the axis, so parity reflects cheaper implementation of similar positions, and at the baseline cost Equal Weight remains ahead, with the abstention path roughly 21 basis points per year behind. The break-even is quoted from the figure, not derived, and it confers no alpha.

Frozen Reverse is the control for that reading. Its turnover is intermediate between the other two paths, yet it stays inferior over the whole displayed range, so low turnover alone cannot rescue a weak direction. Implementation resilience arises only when reduced trading is combined with a path that stays close to the benchmark before costs, which reinforces rather than reverses the main conclusion.

\subsection{Exploratory signed variants}\label{sec:results_exploratory}

The post-holdout signed record is the last real-data exhibit. These variants diagnose whether time-varying orientation and abstention behave more sensibly than a permanent sign, and they are expressly not an independent test of predictive alpha, having been examined after the frozen verdict was known.

\begin{table}
\caption{Exploratory signed variants in the 82-stock universe at the 5-bp baseline cost, ordered by terminal outcome. Annualized log relative performance is reported against Daily Equal Weight and the Matched Fixed Core.}
\label{tab:exploratory_signed_results}
\begin{center}
\footnotesize
\setlength{\tabcolsep}{3.4pt}
\begin{tabular}{lrrrrrrr}
\toprule
Strategy & CAGR & Sharpe & Max.\ DD & Avg.\ turn. & Final wealth & Rel.\ EW & Rel.\ Core \\
\midrule
Clean Consensus Abstention & 14.17\% & 0.815 & $-37.32\%$ & 0.793\% & 3.1233 & $-0.18\%$ & $+0.09\%$ \\
Legacy Cumulative Signed & 14.16\% & 0.815 & $-37.21\%$ & 0.990\% & 3.1216 & $-0.19\%$ & $+0.09\%$ \\
Clean Raw Follow & 14.16\% & 0.814 & $-37.32\%$ & 1.057\% & 3.1210 & $-0.19\%$ & $+0.08\%$ \\
Clean Cumulative Signed & 14.16\% & 0.814 & $-37.32\%$ & 1.057\% & 3.1210 & $-0.19\%$ & $+0.08\%$ \\
Legacy Consensus Abstention & 14.14\% & 0.815 & $-37.21\%$ & 0.852\% & 3.1157 & $-0.21\%$ & $+0.06\%$ \\
Clean Local Signed & 14.10\% & 0.812 & $-37.32\%$ & 1.085\% & 3.1063 & $-0.24\%$ & $+0.03\%$ \\
Legacy Local Signed & 14.06\% & 0.813 & $-37.21\%$ & 1.030\% & 3.0987 & $-0.27\%$ & $+0.00\%$ \\
\midrule
Legacy Raw Follow & 14.04\% & 0.810 & $-37.21\%$ & 0.989\% & 3.0923 & $-0.30\%$ & $-0.02\%$ \\
Legacy Frozen Reverse & 13.88\% & 0.803 & $-38.03\%$ & 0.965\% & 3.0571 & $-0.43\%$ & $-0.16\%$ \\
Clean Frozen Reverse & 13.74\% & 0.798 & $-37.92\%$ & 1.010\% & 3.0252 & $-0.55\%$ & $-0.28\%$ \\
\bottomrule
\end{tabular}
\end{center}
\end{table}

Table~\ref{tab:exploratory_signed_results} orders the record. Consensus abstention leads at the lowest turnover, abstaining on roughly 36\% of decisions in its clean form, every signed variant beats the frozen \Reverse{} paths, the strongest edge past the Matched Fixed Core is a slim $+0.09\%$, and none of the ten paths beats Daily Equal Weight in terminal compounding. Greater sign-switching flexibility does not monotonically help, since the local rules trail the cumulative and consensus ones, and the clean cumulative rule never actually reverses, coinciding with raw \Follow{} throughout. What signed adaptation and abstention demonstrably repair is the handling of a failed orientation, not the absence of a return edge. The demonstrated contribution is behavioral, avoiding forced directional exposure when evidence is weak.

\subsection{Mechanism ablations and synthetic control}\label{sec:results_ablation}

A controller that mostly abstains invites the objection that it is simply too strict, and the real data cannot answer it, since no real signal here was informative. The two exploratory exercises of Section~\ref{sec:baselines} address it, one rerunning the frozen primary path with layers removed, the other running the complete controller on simulated markets whose planted signal has a chosen strength, so engagement can be read as a function of informativeness. Strength is a simulation input, and seed dispersion is simulation dispersion, not a market confidence interval.

\begin{table}
\caption{Layer ablation on the signed controller's reference path in the 82-stock universe, measured against the induced \Safe{} benchmark. Terminal gain and worst shortfall are in basis points of relative wealth. Always \Follow{} and Always \Reverse{} run the fixed action through the same split account; their gains are split-account diagnostics, not alpha. All rows are exploratory.}
\label{tab:ablation_primary}
\begin{center}
\footnotesize
\begin{tabular}{lrrrr}
\toprule
Variant & Terminal gain & Worst shortfall & Mean active prob. & Abstention \\
\midrule
Full controller & $-15.9$ & 15.9 & 0.108 & 62.0\% \\
No gate & $-10.6$ & 10.6 & 0.250 & 0.0\% \\
No budget & $-16.5$ & 16.5 & 0.122 & 61.8\% \\
Neither & $-26.8$ & 26.8 & 0.631 & 0.0\% \\
Always \Safe{} & 0.0 & 0.0 & 0.000 & 100.0\% \\
Always \Follow{} & $+13.4$ & 18.2 & 1.000 & 0.0\% \\
Always \Reverse{} & $-45.9$ & 57.3 & 1.000 & 0.0\% \\
\bottomrule
\end{tabular}
\end{center}
\end{table}

Table~\ref{tab:ablation_primary} reads two ways, and both are reported. The full controller's 15.9-basis-point shortfall sits well below the uncontrolled master's 26.8 and below both fixed orientations, so the safety layers together do limit damage. Yet the budget alone holds the shortfall to 10.6 basis points, below the full controller, because with the gate removed the collateral cap alone keeps executed active mass near one quarter and the master's learned mix then earns slightly more than the gated path on this signal. The gate's contribution on this path is therefore abstention, exposure withheld on 62\% of decisions, and not a smaller shortfall, whereas removing the budget barely changes the outcome because the gate had already removed most of the mass the budget would have capped. Always \Follow{}'s positive terminal gain is the split-account mirror of Table~\ref{tab:fixed_orientation_results}'s finding that \Follow{} beats \Reverse{}, and its 18.2-basis-point worst shortfall shows what the controller's abstention cost in return and bought in protection. The supplement carries the same ablation on the two diagnostic signals, where the ordering of gate and budget reverses for one of them.

\begin{table}
\caption{Synthetic graded response of the complete controller, means over ten seeds of 900 episodes with the seed standard deviation in parentheses. Signal strength is the planted causal contribution to returns, a simulation input. Gains and shortfalls are in basis points of wealth relative to the induced \Safe{} benchmark.}
\label{tab:synthetic_ladder}
\begin{center}
\footnotesize
\begin{tabular}{lrrrr}
\toprule
Strength & Terminal gain & Worst shortfall & Mean active prob. & Abstention \\
\midrule
0.00 & $-0.3$ (0.1) & 0.4 & 0.007 & 97.7\% \\
0.03 & $+26.9$ (1.6) & 0.0 & 0.197 & 35.4\% \\
0.06 & $+63.8$ (2.7) & 0.0 & 0.212 & 33.2\% \\
0.12 & $+145.6$ (5.5) & 0.0 & 0.229 & 32.0\% \\
\bottomrule
\end{tabular}
\end{center}
\end{table}

Table~\ref{tab:synthetic_ladder} shows the response to be graded. With no signal the controller abstains on 98\% of decisions and loses 0.3 basis points to its own residual exploration, which is the behavior the real paths exhibit. At the three positive strengths it funds about a fifth of its probability mass and earns 27, 64, and 146 basis points with a worst shortfall below 0.1 basis points, while Always \Follow{} loses on average at every strength because the planted orientation reverses by regime. Within regimes, the executed mass lands where the regime puts it, \Follow{} mass of 0.26 to 0.30 in \Follow{} regimes, \Reverse{} mass of 0.29 to 0.32 in \Reverse{} regimes, and abstention above 82\% in neutral regimes, with the item-level breakdown in the supplement. The controller engages when the advice is informative and in the direction the advice is informative, and it stays out otherwise.

\section{Discussion and Limitations}\label{sec:discussion}

The results close on two findings, abstention on real data and graded engagement on simulated markets, and the first deserves a more careful weighing than a disappointment reading would give it. Development never established a stable economic direction; it established closeness to a strong benchmark at far lower turnover, which is an implementation property. The frozen test then showed that the directional claim built on top of that property does not transfer, and the polarity diagnosis showed that no fixed sign would have rescued it. The appropriate conclusion is not that the predictor should have been assigned the opposite sign. It is that the mapping from statistical prediction to economically useful portfolio direction is itself uncertain, and a prediction can be informative in some periods without supporting any permanent orientation. Using such a prediction requires separating the generation of advice from the authorization of capital.

\paragraph{Prediction and permission to take risk are different objects.}
Producing a prediction should not automatically create exposure. The controller separates four questions,
\[
\begin{aligned}
&\text{What directional actions are available?}\\
&\text{Which action currently appears economically preferable?}\\
&\text{Is the matured evidence sufficient to justify exposure?}\\
&\text{How much of that exposure can outstanding loss capacity finance?}
\end{aligned}
\]
with the candidate set answering the first, the share-adaptive master the second, the evidence gate the third, and the delayed collateral budget the fourth. The decomposition matters more empirically than any predictor used in the experiments, and delay is why. Under overlapping $H=5$ cohorts several unresolved decisions coexist, so a controller reacting only to the latest prediction creates more directional risk than its matured evidence supports, whatever that prediction's quality. The thesis compresses to one line,
\begin{center}
\principle{prediction $\neq$ permission to risk capital.}
\end{center}

\paragraph{The negative holdout result is part of the contribution.}
A conventional selection workflow could convert the same observation into an apparently successful paper. Observe that \Reverse{} underperforms, flip the sign, pick a favorable local orientation rule, and report the improved path, which is precisely the route by which overfit backtests reach print \citep{BaileyEtAl2016PBO}. The hierarchy instead preserves the failed frozen specification as the cross-universe test and reports the signed rules as mechanism analyses motivated by that failure. Preservation is what makes the negative result informative. It reveals polarity as a quantity that may need to be learned rather than hard-coded, and the exploratory variants support that reading without overturning the verdict, improving materially on the weakest fixed paths while still finishing below Equal Weight. The evidence favors adaptation as damage control and implementation, not as a new source of alpha.

\paragraph{Abstention is an economically meaningful action.}
A portfolio holding only \Follow{} and \Reverse{} must express a directional opinion even when evidence for both is weak. The \Safe{} action changes the structure of the problem, preserving optionality without converting uncertainty into a position, and the realized path shows the option exercised substantively, with the gate repeatedly transferring proposed active mass to \Safe{} and the budget imposing a second reduction when unresolved cohorts had consumed loss capacity. Abstention is a decision, not indecision. It records the gap between evidence enough to rank actions statistically and evidence enough to risk capital economically, and its price is explicit. The penalty $\mathcal{I}_T$ charges the conservative worst case, so removed exposure that would have won is a paid opportunity cost, while removed exposure that would have lost is a saving the bound never credits. A large intervention term may therefore record refused risk rather than poor learning.

\paragraph{Tracking and safety guarantees have different practical roles.}
The tracking branch decomposes performance loss into comparator switching, online learning, delay, numerics, and intervention, and that decomposition, not its numerical value, is its contribution. The bound is deliberately worst-case and can be loose, particularly when the gate removes probability frequently and $\mathcal{I}_T$ dominates; in some configurations it is too conservative to forecast realized performance, which is a property of the question it answers rather than a defect. The safety branch neither needs the prediction to be useful nor the tracking bound to be tight. Its job is limiting damage when learning is hard, and the empirical contrast makes the division visible. The account-level floor permits relative wealth at 80\% of the fallback, while the realized minimum ratio is 0.998406, a 15.9-basis-point worst gap produced by layered conservatism the worst case cannot assume. The practical safety statement is accordingly stronger than the performance statement.

\paragraph{The layers divide the work.}
The ablation and the synthetic ladder together say what each layer is for. In mass removed the gate dominates on the real path, whereas in shortfall the collateral budget alone does better than the full controller, because it caps exposure without regard to evidence, and the gate's contribution is the abstention itself, refusing to fund an orientation the matured record does not support. On simulated markets the controller stays out when the planted signal is absent and otherwise engages at a roughly constant fraction of its mass, with a reward that scales with signal strength and a direction that follows the regime. Neither layer is redundant across the three real signal panels, although on the reference path the budget's reduction is small once the gate has acted, and neither is the whole story. The budget answers how much can be financed, the gate answers whether anything should be, and the realized paths show the two answers can disagree, which is the separation of questions the architecture was built to make visible.

\paragraph{Safety is relative to Safe, not to cash or initial wealth.}
One qualification bounds all of the above. \Safe{} is a conservative fully invested fallback, not a risk-free asset, and it bears its own market drawdowns on the realized path, as the results record. The floor $W_t^{\mathrm{prot}}\geq\rho S_t$ therefore never promises that protected wealth cannot fall in absolute terms. It limits the additional damage attributable to acting on untrusted advice, which is the economically relevant question for an investor with a pre-existing mandate. The paper's use of the word is fixed accordingly,
\begin{equation}
\label{eq:discussion_relative_safety}
\principle{safety $=$ controlled incremental damage relative to a chosen fallback,}
\end{equation}
not guaranteed preservation of nominal wealth.

\paragraph{Transaction-cost resilience is not forecasting alpha.}
The 31-basis-point crossover arises because identical realized positions pay less when trading is cheaper to avoid, with neither signals nor positions recomputed anywhere on the cost axis. Rising frictions do not make the predictor more informative; they make low turnover more valuable, and an execution mechanism can carry that value without improving gross forecasting at all. The reading holds exactly as far as the proportional cost model does, a boundary the limitations below own.

\subsection{Limitations and future evaluation}\label{sec:limitations}

The evidential scope is two universes, and neither can serve again. The 24-stock sample built the architecture and cannot validate it, while the 82-stock universe delivered the frozen verdict, after which every signed variant examined on it is exploratory by construction. The holdout is also cross-sectional rather than temporal, disjoint assets under substantially overlapping calendar history, so it tests asset-universe transfer while holding the macroeconomic environment broadly aligned, and it establishes nothing about behavior in a future regime. Those are different questions, and conflating them would overstate what either universe can show.

The signal side is limited symmetrically. The prediction evidence here is weak, which makes the setting right for studying safe use of unreliable advice and wrong for measuring value preservation under a genuinely informative predictor, so on market data the analysis shows behavior when abstention is frequently appropriate and no more. The only evidence of engagement under an informative signal is the synthetic ladder, whose signal strength is an input the generator sets, so it establishes that the controller can engage, not that any real predictor would make it. The reported paths are realized walk-forward runs, leaving training randomness uncharacterized; a multi-seed characterization of the kind run at scale elsewhere \citep{Ayari2026FeatureConfig} would be needed before conclusions leaned on a neural predictor.

The model scope is bounded by two further choices. The proportional cost model makes the 31-basis-point crossover a sensitivity property of assumed linear frictions and observed turnover paths, not an execution simulation with impact, state-dependent spreads, depth, or capacity. And the architectural constants of Table~\ref{tab:controller_config} delimit the certified domain, so alternative mandates or horizons can change both economic behavior and certificate tightness, with the guarantees requiring realized rewards to stay inside the bounded domain they assume.

Protection is relative to an endogenous fallback, and that is the last boundary. An undesirable \Safe{} policy is not redeemed by protecting a satellite relative to it, and choosing the fallback remains an investment decision outside the learning guarantee, exactly where \eqref{eq:discussion_relative_safety} places it.

One experiment would strengthen the claim, and it is confirmatory rather than more tuning. The complete controller, candidates, master, gate, collateral configuration, protected fraction, and reporting protocol, should be frozen and evaluated on a third untouched universe, ideally with a distinct calendar regime, richer cost and liquidity models, repeated predictor training, and alternative fallback anchors. On the theory side, sharper data-dependent bounds that keep hard collateral safety while charging less for benign abstention would make the tracking certificate informative as a number, not only as a decomposition. The reformulated problem stands either way. When direction and reliability cannot be trusted, producing a better forecast is only part of portfolio learning. The rest is deciding when a forecast deserves capital, how much unresolved predictive risk can be supported, and how to preserve a fallback when the answer is uncertain.

\section{Conclusion}\label{sec:conclusion}

Predictions are abundant; their economic direction, reliability, and timing are not therefore trustworthy. The motivating observation is simple and consequential. A prediction that appears useful in one setting need not retain a stable portfolio orientation when transferred to another asset universe, and a portfolio that hard-codes the orientation inherits that instability as uncontrolled exposure.

The formulation answers by making direction a learned choice inside three safety layers. Prediction-assisted allocation is posed over \Safe{}, \Follow{}, and \Reverse{} actions rather than a permanently fixed sign. A share-adaptive master learns among them from matured Safe-relative rewards, an evidence gate abstains by transferring unsupported active probability to \Safe{}, a delayed collateral budget limits how much directional exposure can be funded while earlier cohorts remain unresolved, and a protected split account restricts the fraction of total wealth the adaptive component can touch.

The theory keeps performance adaptation and safety separate. A tracking branch competes with switching sequences of the three actions while charging delayed decisions and one-sided intervention exactly, and a collateral branch protects pathwise without requiring any prediction to be correct. Realized Safe-relative performance is bounded simultaneously by a comparator-based tracking certificate and an independent hard safety certificate.

The empirics separate the same way. Development showed conservative prediction-assisted portfolios near Daily Equal Weight at a fraction of the turnover, without a return advantage. The frozen 82-stock test then failed to confirm the transferred fixed orientation, and neither fixed sign would have succeeded, the polarity lesson recorded in \eqref{eq:prediction_polarity_failure}. Low-turnover closeness transferred; direction did not.

The mechanisms answered the failure as designed. Exploratory signed and abstention variants repaired much of the damage of the worst fixed paths without a terminal edge over Equal Weight, the gate removed a substantial and time-varying share of proposed active mass, protected wealth stayed within 15.9 basis points of the induced \Safe{} benchmark, the low-turnover abstention variant bought resilience to rising proportional costs, and on simulated markets the controller placed its executed mass on the planted orientation when a signal was present and abstained when none was.

That record justifies a different evaluation standard for prediction-assisted portfolio systems. `Is the prediction accurate?' and `Should this prediction be allowed to risk capital now?' are distinct questions, and a usable controller must answer the second conservatively even while the first is uncertain. The main claim is bounded accordingly. The contribution is a framework for safe adaptive use of untrusted predictions, able to learn their economic polarity, abstain when support is insufficient, account for delayed unresolved exposure, and preserve a chosen fallback when the prediction fails; it is not demonstrated alpha. In compact form,
\begin{center}
\principle{untrusted prediction $\rightarrow$ signed adaptation $\rightarrow$ abstention $\rightarrow$ collateral control $\rightarrow$ protected execution.}
\end{center}

A stronger profitability claim awaits exactly one experiment, the complete controller frozen before evaluation on a third untouched universe. Until then, the evidence supports safe adaptive use of untrusted predictions under delayed feedback, not demonstrated alpha.

\section*{Funding}
The authors received no specific funding for this work.

\section*{Disclosure statement}
No potential competing interest was reported by the authors.

Generative artificial-intelligence tools (Anthropic's Claude) were used in a limited capacity to assist with organizing the literature, grammar checking, improving the coherence and readability of the text, and \LaTeX{} formatting. The research idea, methodology, algorithms, implementation, experiments, and results are entirely the authors' own work. The authors verified all references and reviewed and edited all AI-assisted content, and take full responsibility for the manuscript.

\section*{Data availability statement}
\ifarxiv
The online supplement is appended to this preprint after the references, and the result tables behind every number reported here, each with its SHA-256 hash, accompany the arXiv record as ancillary files. The controller code and the frozen experiment archives are available from the corresponding author on reasonable request. The underlying daily market and accounting data were obtained from commercial vendors under license and cannot be redistributed; the feature definitions required to reconstruct the panels are described in Section~\ref{sec:universes} and the supplement.
\fi

\bibliographystyle{rQUF}
\bibliography{references}

\appendix
\setlength{\abovedisplayskip}{5pt plus 1pt minus 1pt}
\setlength{\belowdisplayskip}{5pt plus 1pt minus 1pt}
\setlength{\abovedisplayshortskip}{3pt plus 1pt}
\setlength{\belowdisplayshortskip}{3pt plus 1pt}
\renewcommand{\theequation}{A.\arabic{equation}}
\section{Proofs}\label{app:proofs}

This appendix makes the manuscript self-contained. Six supporting results are stated and proved first, then every theorem, lemma, and remark of Section~\ref{sec:theory} whose proof the text deferred is proved in statement order, all in the notation of Sections~\ref{sec:methodology} and \ref{sec:theory}, unchanged. Appendix~\ref{app:pseudocode} records the controller pseudocode and the collateral bookkeeping. The online supplement carries no theorems or proofs; it holds the remaining controller specification, the implementation and data audits, and additional empirical diagnostics.

\subsection{Supporting results}\label{app:supporting}

\begin{lemma}[Exact split-account aggregation]\label{lem:aggregation}
Consider a cohort with initial wealth $V>0$, candidate probabilities $p\in\Delta_A$, and positive candidate net wealth factors $F_1,\ldots,F_A$. If initial capital $p_aV$ is assigned to candidate subaccount $a$, the aggregate terminal wealth is $V\sum_a p_aF_a$, so the aggregate cohort wealth factor is exactly $F^{\mathrm{mix}}=\sum_a p_aF_a$.
\end{lemma}

\begin{proof}
Subaccount $a$ begins with wealth $p_aV$ and, by definition of its factor, ends with $p_aVF_a$. Summing over subaccounts and dividing by $V>0$ gives the claim. The identity is exact because capital aggregates at the subaccount level; nothing is approximated at the level of log returns.
\end{proof}

The lemma is what the split-account identity of (A6) asserts operationally, and the implementation audits verify it on the realized path.

\begin{lemma}[Log-concavity dominance]\label{lem:logconcavity}
With action $0$ denoting \Safe{}, $F_0>0$, $\ell_a=\log(F_a/F_0)$, and $x_a=\ell_a/B$, every $p\in\Delta_A$ satisfies
$\log\bigl(\sum_a p_aF_a/F_0\bigr)\geq\sum_a p_a\ell_a=Bp^{\top}x$.
\end{lemma}

\begin{proof}
Write $\sum_a p_aF_a/F_0=\sum_a p_a(F_a/F_0)$ and apply Jensen's inequality to the concave logarithm, giving $\log\sum_a p_a(F_a/F_0)\geq\sum_a p_a\log(F_a/F_0)=\sum_a p_a\ell_a$, which equals $Bp^{\top}x$ by the normalization. The slack is generally positive, so the linear reward $p^{\top}x$ used by the learner understates the realized mixed cohort.
\end{proof}

\begin{lemma}[Fixed-Share path-mixture representation]\label{lem:pathmixture}
For share rate $\gamma$, define the path prior $\Pi_\gamma(a_{1:T})=\pi_{a_1}\prod_{t=2}^{T}P_\gamma(a_t\mid a_{t-1})$, so that $\mathcal{C}_\gamma=-\log\Pi_\gamma$. The Fixed-Share recursion \eqref{eq:fixed_share} produces exactly the one-period marginals of the exponentially weighted distribution over action paths with prior $\Pi_\gamma$, and its partition function satisfies
$Z_{t+1}/Z_t=\sum_{j}q^{(\gamma)}_{t,j}\,e^{\eta x_{t,j}}$.
\end{lemma}

\begin{proof}
Assign each partial path $a_{1:t}$ the pre-reward weight $W_t(a_{1:t})=\pi_{a_1}\bigl[\prod_{u=2}^{t}P_\gamma(a_u\mid a_{u-1})\bigr]\exp\bigl(\eta\sum_{u=1}^{t-1}x_{u,a_u}\bigr)$ and let $Z_t=\sum_{a_{1:t}}W_t(a_{1:t})$, with $Z_1=\sum_{a_1}\pi_{a_1}=1$. The marginal probability of action $j$ at round $t$ is $q^{(\gamma)}_{t,j}=Z_t^{-1}\sum_{a_{1:t}:a_t=j}W_t(a_{1:t})$. Observing $x_t$ multiplies the weight of paths ending in $j$ by $e^{\eta x_{t,j}}$, and propagation through $P_\gamma(\cdot\mid j)$, whose rows sum to one, produces the next round's weights. The induced marginal update is exponential reward weighting followed by the share transition, which is \eqref{eq:fixed_share}, and summing the reweighted terms gives the stated partition ratio.
\end{proof}

\begin{proposition}[Per-base Fixed-Share tracking]\label{prop:perbase}
For every share rate $\gamma$, bounded reward sequence, and comparator path $a_{1:T}$,
\[
\sum_{t=1}^{T}q_t^{(\gamma)\top}x_t
\;\geq\;
\sum_{t=1}^{T}x_{t,a_t}-\frac{\mathcal{C}_\gamma(a_{1:T})}{\eta}-\frac{\eta T}{2}.
\]
\end{proposition}

\begin{proof}
By Lemma~\ref{lem:pathmixture}, $\log Z_{T+1}=\sum_{t=1}^{T}\log\bigl(\sum_j q^{(\gamma)}_{t,j}e^{\eta x_{t,j}}\bigr)$. Hoeffding's lemma for a variable supported on $[-1,1]$ gives $\log\mathbb{E}\,e^{\eta X}\leq\eta\,\mathbb{E}[X]+\eta^{2}/2$, so each term is at most $\eta\,q_t^{(\gamma)\top}x_t+\eta^{2}/2$ and $\log Z_{T+1}\leq\eta\sum_t q_t^{(\gamma)\top}x_t+\eta^{2}T/2$. The partition function also dominates the contribution of the single comparator path, $Z_{T+1}\geq\Pi_\gamma(a_{1:T})\exp\bigl(\eta\sum_t x_{t,a_t}\bigr)$, so $\log Z_{T+1}\geq-\mathcal{C}_\gamma(a_{1:T})+\eta\sum_t x_{t,a_t}$. Combining the two bounds and dividing by $\eta>0$ proves the claim.
\end{proof}

\begin{proposition}[Geometric share-grid complexity]\label{prop:gridcomplexity}
Let $n=T-1$ and let $K$ count the switches of $a_{1:T}$. The grid $\Gamma$ of \eqref{eq:share_grid} with $\gamma_{\max}=\tfrac12$ satisfies $\mathcal{C}_0(a_{1:T})=-\log\pi_{a_1}$ when $K=0$, and for every $1\leq K\leq n$,
\begin{equation}
\label{eq:grid_complexity_bound}
\min_{1\leq m\leq M}\mathcal{C}_{\gamma_m}(a_{1:T})
\;\leq\;
-\log\pi_{a_1}+K\log\Bigl(\frac{2eA(T-1)}{K}\Bigr).
\end{equation}
\end{proposition}

\begin{proof}
A path with $K$ switches has exact complexity $\mathcal{C}_\gamma=-\log\pi_{a_1}-(n-K)\log\bigl(1-\gamma+\gamma/A\bigr)-K\log(\gamma/A)$. For $K=0$ take $\gamma=0$, under which every stay has probability one. For $1\leq K\leq n/2$, the geometric grid contains a rate with $K/(2n)\leq\gamma\leq K/n$, since consecutive grid points differ by a factor of two below $\gamma_{\max}$. Bounding $1-\gamma+\gamma/A\geq1-\gamma$ gives $\mathcal{C}_\gamma\leq-\log\pi_{a_1}+(n-K)\log\bigl(1/(1-\gamma)\bigr)+K\log(A/\gamma)$. The stay term obeys $(n-K)\log\bigl(1/(1-\gamma)\bigr)\leq(n-K)\log\bigl(n/(n-K)\bigr)=(n-K)\log\bigl(1+K/(n-K)\bigr)\leq K$ using $\gamma\leq K/n$, and the switch term obeys $K\log(A/\gamma)\leq K\log(2An/K)$ using $\gamma\geq K/(2n)$; together they give \eqref{eq:grid_complexity_bound}. For $K>n/2$ take $\gamma_{\max}=1/2$, under which $1-\gamma+\gamma/A\geq1/2$ and $\gamma/A=1/(2A)$, so $\mathcal{C}_{1/2}\leq-\log\pi_{a_1}+(n-K)\log2+K\log(2A)\leq-\log\pi_{a_1}+K\log(4A)$, and $\log(4A)\leq\log(2eAn/K)$ because $n/K\geq1$ and $2e>4$.
\end{proof}

\begin{proposition}[Outer-Hedge regret]\label{prop:outerhedge}
Let $y_{t,m}=q_t^{(m)\top}x_t\in[-1,1]$ be the reward of base $m$ and let $\bar{q}_t=\sum_m v_{t,m}q_t^{(m)}$ under the Hedge update of \eqref{eq:share_grid} with uniform prior. For every base $m$,
\[
\sum_{t=1}^{T}\bar{q}_t^{\top}x_t
\;\geq\;
\sum_{t=1}^{T}y_{t,m}-\frac{\log M}{\beta}-\frac{\beta T}{2}.
\]
\end{proposition}

\begin{proof}
Linearity gives $\bar{q}_t^{\top}x_t=\sum_m v_{t,m}y_{t,m}$. With the potential $U_t=\sum_m v_{1,m}\exp\bigl(\beta\sum_{u<t}y_{u,m}\bigr)$, the update implies $U_{t+1}/U_t=\sum_m v_{t,m}e^{\beta y_{t,m}}$, and Hoeffding's lemma bounds each ratio's logarithm by $\beta\sum_m v_{t,m}y_{t,m}+\beta^{2}/2$, so $\log U_{T+1}\leq\beta\sum_t\bar{q}_t^{\top}x_t+\beta^{2}T/2$. For any fixed base, $U_{T+1}\geq v_{1,m}\exp\bigl(\beta\sum_t y_{t,m}\bigr)$, so $\log U_{T+1}\geq-\log M+\beta\sum_t y_{t,m}$ under the uniform prior. Combining and dividing by $\beta>0$ proves the claim.
\end{proof}

\subsection{Proof of Theorem~\ref{thm:share_adaptive_tracking}}

\begin{proof}
For every base $m$, Proposition~\ref{prop:perbase} bounds its cumulative reward, $\sum_t y_{t,m}\geq\sum_t x_{t,a_t}-\mathcal{C}_{\gamma_m}(a_{1:T})/\eta-\eta T/2$. Substituting into Proposition~\ref{prop:outerhedge} gives, for every $m$,
\[
\sum_{t=1}^{T}\bar{q}_t^{\top}x_t
\;\geq\;
\sum_{t=1}^{T}x_{t,a_t}-\frac{\mathcal{C}_{\gamma_m}(a_{1:T})}{\eta}-\frac{\eta T}{2}-\frac{\log M}{\beta}-\frac{\beta T}{2},
\]
and taking the best base yields the minimum over the grid in \eqref{eq:share_adaptive_tracking_bound}. The explicit forms follow from Proposition~\ref{prop:gridcomplexity}. A static comparator pays only $-\log\pi_{a_1}$, and a comparator with $K$ switches pays at most $-\log\pi_{a_1}+K\log(2eA(T-1)/K)$, which is the $O(K\log(AT/K))$ price quoted after the theorem. All bases share one $\eta$ and differ only in $\gamma$, so the outer layer adapts the assumed switching intensity, exactly as the name of the theorem asserts.
\end{proof}

\subsection{Proof of Theorem~\ref{thm:exact_delay}}

\begin{proof}
Fix a cohort $s$. The reward gap between the decision-time and immediate-feedback states is an inner product with the master-state gap,
\[
\bigl(q_s^{\mathrm{dec}}\bigr)^{\top}x_s-\bar{q}_s^{\top}x_s
=\bigl(q_s^{\mathrm{dec}}-\bar{q}_s\bigr)^{\top}x_s
\;\geq\;
-\bigl\|q_s^{\mathrm{dec}}-\bar{q}_s\bigr\|_1\,\bigl\|x_s\bigr\|_{\infty}
\;\geq\;
-\bigl\|\bar{q}_{\kappa(s)}-\bar{q}_s\bigr\|_1,
\]
where the first inequality is H\"older's and the second uses $\|x_s\|_{\infty}\leq1$ from (A1) together with $q_s^{\mathrm{dec}}=\bar{q}_{\kappa(s)}$ from (A3). Summing over $s=1,\ldots,T$ collects the per-cohort gaps into the realized penalty,
\[
\sum_{s=1}^{T}\bigl(q_s^{\mathrm{dec}}\bigr)^{\top}x_s
\;\geq\;
\sum_{s=1}^{T}\bar{q}_s^{\top}x_s
-\sum_{s=1}^{T}\bigl\|\bar{q}_s-\bar{q}_{\kappa(s)}\bigr\|_1
=\sum_{s=1}^{T}\bar{q}_s^{\top}x_s-\Delta_T^{\mathrm{exact}},
\]
which is \eqref{eq:exact_delay_reduction}. No step introduces slack beyond the H\"older bound, and every quantity on the right is measured from the recorded master path, so the penalty is realized rather than worst-case over delay processes.
\end{proof}

\subsection{Proof of Lemma~\ref{lem:one_sided_intervention}}

\begin{proof}
By (A4), for every active action $a\neq\Safe$,
\[
0
\leq
q^{\mathrm{exec}}_{s,a}
\leq
q^{\mathrm{gate}}_{s,a}
\leq
\widehat q_{s,a}.
\]
The first intervention either preserves an active coordinate or
transfers it to \Safe{}, while the collateral budget can only apply a
common factor in $[0,1]$ to the surviving active coordinates. Hence
neither stage can increase active probability.

Define
\[
\delta_{s,a}
=
\widehat q_{s,a}
-
q^{\mathrm{exec}}_{s,a}
\geq0,
\qquad
a\neq\Safe,
\]
so that
\[
I_s
=
\sum_{a\neq\Safe}\delta_{s,a}.
\]
Because both probability vectors lie in the simplex and all removed
active mass is transferred to \Safe{},
\[
q^{\mathrm{exec}}_{s,\Safe}
-
\widehat q_{s,\Safe}
=
\sum_{a\neq\Safe}
\left(
\widehat q_{s,a}
-
q^{\mathrm{exec}}_{s,a}
\right)
=
I_s.
\]
Consequently,
\begin{align}
\left\|
q_s^{\mathrm{exec}}
-
\widehat q_s
\right\|_1
&=
\sum_{a\neq\Safe}
\left(
\widehat q_{s,a}
-
q^{\mathrm{exec}}_{s,a}
\right)
+
\left(
q^{\mathrm{exec}}_{s,\Safe}
-
\widehat q_{s,\Safe}
\right)
\nonumber\\
&=
2I_s,
\end{align}
which proves
\eqref{eq:one_sided_l1_identity}.

For the reward comparison, $x_{s,\Safe}=0$ by (A1), and therefore
\begin{align}
\bigl(q_s^{\mathrm{exec}}\bigr)^{\top}x_s
-
\widehat q_s^{\top}x_s
&=
-\sum_{a\neq\Safe}
\left(
\widehat q_{s,a}
-
q^{\mathrm{exec}}_{s,a}
\right)
x_{s,a}
\nonumber\\
&\geq
-\sum_{a\neq\Safe}
\left(
\widehat q_{s,a}
-
q^{\mathrm{exec}}_{s,a}
\right)
\nonumber\\
&=
-I_s,
\end{align}
where the inequality uses
$x_{s,a}\leq1$
and
$\widehat q_{s,a}-q^{\mathrm{exec}}_{s,a}\geq0$.
Thus
\[
\bigl(q_s^{\mathrm{exec}}\bigr)^{\top}x_s
\geq
\widehat q_s^{\top}x_s-I_s.
\]
Summing over $s=1,\ldots,T$ gives
\eqref{eq:cumulative_one_sided_reward_bound}.
\end{proof}

\subsection{Proof of Theorem~\ref{thm:anytime_collateral}}

\begin{proof}
The collateral account is the state $(\mathsf{R}_\tau,\mathsf{P}_\tau,G_\tau,J_\tau,D_\tau)$, and it changes only at openings and settlements. Before the first event, $\mathsf{P}_0=G_0=J_0=D_0=0$ and $\mathsf{R}_0\geq0$ by the initial solvency of (A5), so solvency $\mathsf{R}_\tau\geq\mathsf{P}_\tau$ and the identity $\mathsf{R}_\tau=\mathsf{R}_0+G_\tau+J_\tau-D_\tau$ both hold at $\tau=0$. The proof is an induction over events, showing that each event type preserves both.

At the opening of cohort $s$, the commitment satisfies $c_s\leq b_s^{\mathrm{avail}}=\mathsf{R}_\tau-\mathsf{P}_\tau$ by (A5), and the only change of state is $\mathsf{P}_{\tau+1}=\mathsf{P}_\tau+c_s$. Hence
\[
\mathsf{R}_{\tau+1}-\mathsf{P}_{\tau+1}
=
\mathsf{R}_\tau-\mathsf{P}_\tau-c_s
\geq0,
\]
and the identity is unchanged because $\mathsf{R}$, $G$, $J$, and $\Omega$ are unchanged.

At the settlement of cohort $s$, the realized reward $g_s=(q_s^{\mathrm{exec}})^{\top}x_s$ of the probability stored at opening satisfies $g_s\geq-c_s$ by (A1) and (A5), the commitment is released, a replenishment $\xi_s\geq0$ may be applied, and a withdrawal $\omega_s\geq0$ may be banked. The state moves to
\[
\mathsf{R}_{\tau+1}=\mathsf{R}_\tau+g_s+\xi_s-\omega_s,
\quad
\mathsf{P}_{\tau+1}=\mathsf{P}_\tau-c_s,
\quad
G_{\tau+1}=G_\tau+g_s,
\quad
J_{\tau+1}=J_\tau+\xi_s,
\quad
\Omega_{\tau+1}=\Omega_\tau+\omega_s,
\]
so the identity is preserved term by term. Before the withdrawal the solvency margin is
\[
\mathsf{R}_\tau+g_s+\xi_s-\mathsf{P}_{\tau+1}
=
(\mathsf{R}_\tau-\mathsf{P}_\tau)+(g_s+c_s)+\xi_s
\geq0,
\]
since each bracket is nonnegative by the induction hypothesis and by $g_s\geq-c_s$, and $\xi_s\geq0$. By (A5) a withdrawal is admissible only if the outstanding commitments remain collateralized, that is $\mathsf{R}_{\tau+1}\geq\mathsf{P}_{\tau+1}$ after $\omega_s$ is deducted, so solvency survives the settlement. Induction over all events gives $\mathsf{R}_\tau\geq\mathsf{P}_\tau$ and the accounting identity at every $\tau$.

Substituting the identity into the solvency margin,
\[
G_\tau-\mathsf{P}_\tau
=
(\mathsf{R}_\tau-\mathsf{P}_\tau)-\mathsf{R}_0-J_\tau+\Omega_\tau
\geq
-(\mathsf{R}_0+J_\tau),
\]
because $\mathsf{R}_\tau-\mathsf{P}_\tau\geq0$ and $\Omega_\tau\geq0$. Multiplying by $B>0$ gives \eqref{eq:anytime_completion_certificate}.
\end{proof}

\subsection{Proof of Remark~\ref{rem:protected_floor}}

\begin{proof}
On every admissible date, $W_t^{\mathrm{prot}}=\rho S_t+V_t$ with $S_t>0$ and $V_t\geq0$ by (A6). Dividing by $S_t$,
\[
\frac{W_t^{\mathrm{prot}}}{S_t}=\rho+\frac{V_t}{S_t}\geq\rho,
\]
which is $W_t^{\mathrm{prot}}\geq\rho S_t$. The bound uses no property of the prediction, the master, the gate, or the realized active rewards.
\end{proof}

\renewcommand{\theequation}{B.\arabic{equation}}
\setcounter{equation}{0}
\section{Controller pseudocode and collateral bookkeeping}\label{app:pseudocode}

Algorithm~\ref{alg:controller} states the decision cycle of Section~\ref{sec:protected_execution} as pseudocode on the two clocks of Figure~\ref{fig:architecture}; the collateral rules that (A5) abstracts follow it, and the remaining specification is in the online supplement.

\begin{algorithm}[H]
\small
\caption{Causal signed portfolio control with delayed feedback}
\label{alg:controller}
\begin{algorithmic}[1]
\Require candidate construction rule, horizon $H$, reward bound $B$, signed master, evidence gate, collateral budget
\For{each eligible trading date $t$}
  \If{a new cohort $s$ can be opened}
    \State construct $w_s^{\Safe},w_s^{\Follow},w_s^{\Reverse}$ from pre-return information
    \State $q_s^{\mathrm{raw}}\gets\textsc{MasterProbability}()$
    \State $q_s^{\mathrm{gate}}\gets\textsc{EvidenceGate}(q_s^{\mathrm{raw}})$
    \State $q_s^{\mathrm{exec}}\gets\textsc{CollateralAllocate}(q_s^{\mathrm{gate}})$ \Comment{reserves commitment $c_s$}
    \State $w_s^{\mathrm{exec}}\gets\sum_a q^{\mathrm{exec}}_{s,a}w_s^{a}$; register maturity $m_s=t+H-1$
  \EndIf
  \State observe the current return; mark forward all open \Safe{}, adaptive, and shadow accounts
  \If{an older cohort $j$ matures after the current return}
    \State compute $F_{j,a}$, $\ell_{j,a}=\log(F_{j,a}/F_{j,\Safe})$, verify $|\ell_{j,a}|\leq B$, set $x_{j,a}=\ell_{j,a}/B$
    \State $\textsc{CollateralSettle}(q_j^{\mathrm{exec}},x_j)$ \Comment{uses the probability stored at opening}
    \State $\textsc{MasterUpdate}(x_j)$; \ $\textsc{GateUpdate}(x_j)$
  \EndIf
\EndFor
\end{algorithmic}
\end{algorithm}

Collateral is held in normalized units, one unit of active probability against one unit of worst-case bounded reward. The account maintains the reserve $\mathsf{R}$, the per-cohort commitments $c_j$ of all open cohorts with total $\mathsf{P}$, and the unencumbered budget $b_s^{\mathrm{avail}}=\mathsf{R}-\mathsf{P}$ quoted to each opening. Authorization scales the gated active coordinates by the common factor $\zeta_s=\min\{a_s^{\mathrm{gate}},b_s^{\mathrm{avail}}\}/a_s^{\mathrm{gate}}$, records $c_s=a_s^{\mathrm{exec}}$ before capital is exposed, and holds the commitment until maturity. Settlement credits the realized reward $g_s=(q_s^{\mathrm{exec}})^{\top}x_s$ of the stored opening probability, releases $c_s$, applies any scheduled replenishment $\xi_s\geq0$, distributed evenly across the expected number of matured updates, and banks any excess of the reserve above its operating cap as withdrawn gain $\omega_s\geq0$, admissible only when the remaining commitments stay covered. The implementation checks the solvency margin $\mathsf{R}-\mathsf{P}\geq0$ and the accounting identity $\mathsf{R}=\mathsf{R}_0+G+J-\Omega$ at every event, and verifies $g_s\geq-c_s$ at every settlement, which is how (A5) is audited.

\ifarxiv
\clearpage
\setcounter{section}{0}\setcounter{subsection}{0}
\setcounter{equation}{0}\setcounter{figure}{0}\setcounter{table}{0}
\renewcommand{\thesection}{S\arabic{section}}
\renewcommand{\thesubsection}{\thesection.\arabic{subsection}}
\renewcommand{\theequation}{S.\arabic{equation}}
\renewcommand{\thefigure}{S.\arabic{figure}}
\renewcommand{\thetable}{S.\arabic{table}}
\makeatletter
\def\cl@section{\@elt{subsection}}
\let\@seccntformat\@sseccntformat
\makeatother
\section*{Online supplement}
\section{Scope and conventions}\label{s:scope}

This supplement accompanies the manuscript and is not self-contained. It holds four parts, in dependency order. Section~\ref{s:spec} restores the mechanism-by-mechanism controller specification that the manuscript's Methodology presents at design level, for the parts the manuscript does not itself record; the decision-cycle pseudocode and the complete collateral bookkeeping are in the manuscript's Appendix~B. Section~\ref{s:audits} details the implementation and data audits the manuscript summarizes in single sentences. Section~\ref{s:diagnostics} reports additional empirical diagnostics on the frozen path, one figure and three tables of item-level audit evidence. Section~\ref{s:ablation} gives the layer ablation on all three real signal panels and the regime-level placement behind the manuscript's synthetic ladder. Every mathematical result and proof lives in the manuscript, in its Theoretical Guarantees section and Appendix~A; the supplement states nothing and proves nothing.

Supplement sections, equations, figures, and tables are numbered with an S prefix. Notation is carried unchanged from the manuscript's Methodology and Theoretical Guarantees sections; the only objects defined here and not in the manuscript are the cross-sectional prediction $z_t$ that the candidates are built from, the gate's sample standard deviation, and the development uncertainty radius, which the frozen controller does not use. Manuscript equations, tables, and figures are cited by name rather than number, since the supplement is published as submitted while the manuscript may be renumbered in production.

\section{Complete controller specification}\label{s:spec}

The parts below map onto the manuscript's Methodology subsections. The setting and candidates are specified in full in \ref{s:candidates}; the share-adaptive master in \ref{s:master}, including the share-grid construction and the matured-reward processing order; the evidence gate in \ref{s:gate}; and the split account and event cycle in \ref{s:split}. The complete decision loop is Algorithm~1 of the manuscript's Appendix~B, which also carries the collateral bookkeeping, and the learning-rate rule is stated with the controller configuration in the manuscript's Experimental Design.

\subsection{Candidates}\label{s:candidates}

The cross-sectional predictive signal $z_t$ is mapped to a centered directional vector $d_t$ with $\1^{\top}d_t=0$ and $\|d_t\|_1=1$, so overlays redistribute mass without changing total investment; a zero signal yields a zero direction and both active candidates coincide with \Safe{}. The \Safe{} anchor is the OPMD target defined in the manuscript's Methodology, a Kullback--Leibler proximal step from the drifted holdings with an $\ell_1$ movement penalty, applied to the conservative gradient $g_t^{\mathrm{safe}}$. The active candidates request the overlay $\delta=0.10$ in either orientation, and the feasibility scales $\alpha_t^{\pm}$ are computed as the largest coefficients in $[0,1]$ keeping $w_t^{\Safe}\pm\alpha\delta_t d_t$ inside the capped simplex with $w_{\max}=0.20$; when a coordinate binds, the scale contracts the whole overlay rather than clipping single coordinates, preserving the centered direction. On the audited 82-stock path the full overlay is feasible throughout, $\alpha_t^{+}=\alpha_t^{-}=1$ up to numerical precision (Table~\ref{tab:candidate_audit}). Any probability mixture of the three candidates stays feasible by convexity of the simplex.

\subsection{Master mechanics}\label{s:master}

The action prior is Safe-biased, $\pi=(\pi_{\Safe},(1-\pi_{\Safe})/2,(1-\pi_{\Safe})/2)$, splitting the non-Safe mass equally between orientations, and every Fixed-Share base starts at $q^{(\gamma)}[0]=\pi$. The shared action learning rate and the outer rate follow the closed forms stated with the controller configuration in the manuscript, which give $\eta=0.031879$ and $\beta=0.048711$ at $T=2{,}162$. The share grid is
\[
\Gamma=\operatorname{sortunique}\Bigl(\{0,\gamma_{\max}\}\cup\Bigl\{\tfrac{2^{j}}{T-1}:\tfrac{2^{j}}{T-1}<\gamma_{\max},\ j=0,1,\ldots\Bigr\}\Bigr),
\]
with $\gamma_{\max}=0.50$, giving $M=13$ bases. When a cohort matures, the processing order inside \textsc{MasterUpdate} is fixed. Each base's expected reward $y_{k,m}=(q^{(\gamma_m)}[k-1])^{\top}x_k$ is computed from its pre-update state, the outer Hedge weights are updated from those rewards, and only then does every base apply the exponential reward update and its share transition. The raw decision probability is the state indexed by matured-reward count, $q_s^{\mathrm{raw}}=\bar{q}[\kappa(s)]$, never by calendar count.

\subsection{Evidence gate}\label{s:gate}

For each active action, the gate maintains the evidence summary $\mathcal{S}_{k,a}=(n_{k,a},\widehat{\mu}_{k,a},\underline{\mu}_{k,a})$ over fully matured bounded rewards only. The summary is a rolling window of the last 50 matured rewards of that action, $n_{k,a}\leq50$, with $\widehat{\mu}_{k,a}$ their sample mean and $\underline{\mu}_{k,a}=\widehat{\mu}_{k,a}-1.645\,s_{k,a}/\sqrt{n_{k,a}}$ the one-sided normal lower bound at the 5\% level from the sample standard deviation $s_{k,a}$. The action is accepted when $n_{k,a}\geq25$ and $\underline{\mu}_{k,a}>0$. Consecutive matured rewards overlap through the $H=5$ cohort structure and the standard error does not correct for that dependence, so the level is nominal and the gate is a calibrated screen, not a sequential false-positive guarantee; the safety certificates do not depend on it. The acceptance indicators act multiplicatively on the raw active coordinates and the removed mass is added to the \Safe{} coordinate, so the four execution states are both accepted, one accepted, or full abstention at the unit mass on \Safe{}. The gate state obeys the same maturity clock as the master; no current or partially realized cohort influences an opening decision.

\subsection{Split account, satellite, and event cycle}\label{s:split}

The \Safe{} benchmark runs as $H$ sleeves with $S_{0,h}=1/H$, each implementing the \Safe{} target on its own cohort timing and paying the same proportional costs as every other account; the protected core is its $\rho$-scaled path with $\rho=0.80$. The satellite runs the same sleeve structure with $V_{0,h}=(1-\rho)/H$, as in the manuscript's protected-core display. When sleeve $h$ is refreshed by cohort $s$ it trades to $w_s^{\mathrm{exec}}$, and its shadow accounts carry all three candidates to maturity so that the exact factor identity $F_s^{\mathrm{exec}}=\sum_a q^{\mathrm{exec}}_{s,a}F_{s,a}$ can be verified rather than assumed. The daily cycle is the manuscript's two-phase order. Opening uses only pre-return information, in the sequence candidates, master query, gate, budget, cohort registration; the return is then realized and every open sleeve marked forward; same-date maturities settle last, in the order budget, master, gate. A cohort maturing on date $t$ therefore cannot influence the cohort opened earlier on date $t$.

\section{Implementation and data audits}\label{s:audits}

The audits are implementation-verification exercises, never tests of investment alpha, and they come in three families, data alignment across universes, development-stage machinery, and controller invariants on realized paths.

The data-alignment audit enforces the identity of the 40-feature dictionary across the two universes item by item, 17 price-derived and 23 technical and accounting definitions in fixed order, with no feature added to or removed from the 82-stock panel relative to the 24-stock panel. Accounting variables are verified to enter only in lagged form, so no value can affect a decision before it is economically observable, and the two daily panels are calendar-aligned under the same construction code.

The development uncertainty radius belongs to the development pipeline only, and the frozen controller does not use it. For each validation date $u$ in the walk-forward validation block $\mathcal{V}_t$, the residual-advice error is $e_u=\|\Delta g_u-\widehat{\Delta g}_u\|_{\infty}$, the gap between the realized and predicted change of the development target, and the radius is the empirical quantile $\epsilon_t=Q_{0.75}(\{e_u:u\in\mathcal{V}_t\})$ over the most recent 63-day block. It calibrates the development-stage conservative execution rules compared in the manuscript's development-universe results table.

The controller audits tie the realized code path to the analyzed controller. Candidate feasibility and simplex preservation hold with the full requested overlay on every audited episode (Table~\ref{tab:candidate_audit}). The gate and budget were probed over 24 states drawn from six gate instances and checked on eight realized trajectory tables; no probe or trajectory increased active probability, the Safe-transfer identity and the common budget-scaling property held with zero numerical error, and the minimum audited intervention-certificate margin, the smallest slack by which the manuscript's one-sided intervention inequality held across the probed states and trajectory tables in normalized reward units, was $0.002176$ (Table~\ref{tab:intervention_audit}). Reward boundedness $|\ell_{s,a}|\leq B$ was verified at every maturity, with the largest observed $|\ell_{s,a}|$ on the reference path at 0.0081 against $B=0.05$; the delayed chronology was checked against the event ledger with no opening influenced by a same-date maturity; the split-account aggregation identity held to $2\times10^{-16}$; and the daily protected-wealth invariant was evaluated on four audited paths (Table~\ref{tab:protected_paths}).

\section{Additional empirical diagnostics}\label{s:diagnostics}

The exhibits below complement the manuscript's results figures with finer instrumentation on the same frozen path, each tied to the audit family it visualizes.

\begin{figure}[t]
\centering
\includegraphics[width=0.95\textwidth]{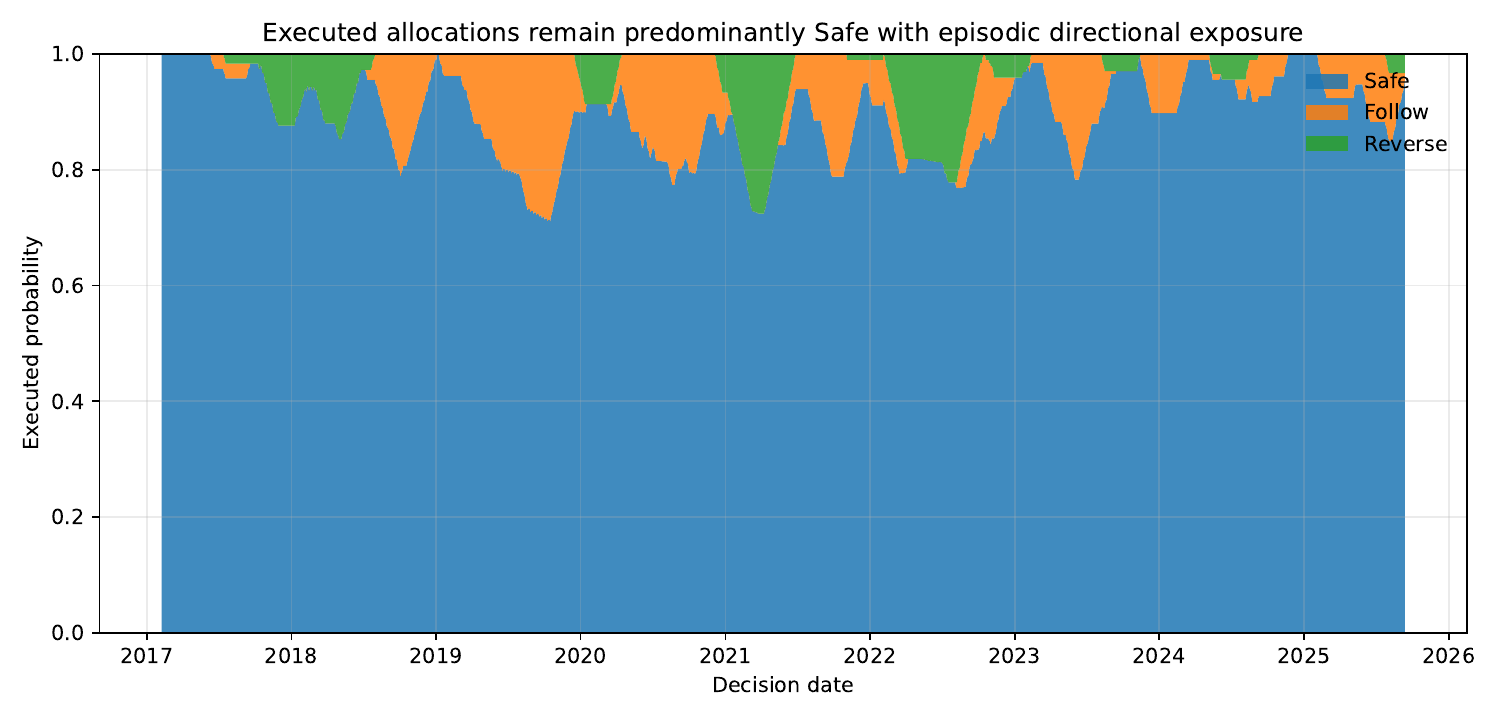}
\caption{Executed allocations remain predominantly \Safe{} with episodic directional exposure. Stacked final probabilities assigned to \Safe{}, \Follow{}, and \Reverse{} after all interventions, as trailing 63-decision rolling means over opened cohorts. An implementation diagnostic of the post-holdout signed controller, not a confirmatory return comparison.}
\label{fig:allocations}
\end{figure}

Figure~\ref{fig:allocations} shows the complete executed probability path. Allocation shifts coincide with gate rejections and budget-binding episodes, matching the intervention decomposition of the manuscript's safety-intervention figure, and the funded portfolio is far more conservative than the raw master because unsupported or uncollateralized active probability is transferred to \Safe{} before capital is committed.

\begin{table}[t]
\caption{Candidate-feasibility audit in the 82-stock universe. \Follow{} and \Reverse{} are constructed symmetrically around the \Safe{} candidate.}
\label{tab:candidate_audit}
\begin{center}
\small
\begin{tabular}{lrr}
\toprule
Diagnostic & Clean full signal & Legacy full signal \\
\midrule
Completed cohorts & 2{,}162 & 2{,}162 \\
Zero-direction rate & 0 & 0 \\
Minimum \Follow{} feasibility scale & 1.000 & 1.000 \\
Minimum \Reverse{} feasibility scale & 1.000 & 1.000 \\
Mean $\ell_1$ distance, \Follow{} vs.\ \Safe{} & 0.1000 & 0.1000 \\
Mean $\ell_1$ distance, \Reverse{} vs.\ \Safe{} & 0.1000 & 0.1000 \\
Minimum candidate weight & 0.000694 & 0.000694 \\
Maximum candidate weight & 0.030928 & 0.030928 \\
Maximum simplex-sum error & $\approx10^{-12}$ & $\approx10^{-12}$ \\
\bottomrule
\end{tabular}
\end{center}
\end{table}

\begin{table}[t]
\caption{Implementation audit of the evidence-gate and collateral-budget maps.}
\label{tab:intervention_audit}
\begin{center}
\small
\begin{tabular}{lr}
\toprule
Audit quantity & Realized value \\
\midrule
Gate instances probed & 6 \\
Gate probe states & 24 \\
Trajectory tables audited & 8 \\
Maximum active-probability increase in gate probes & 0 \\
Maximum active-probability increase on realized trajectories & 0 \\
Maximum Safe-transfer identity error & 0 \\
Maximum common budget-scaling error & 0 \\
Minimum total intervention-certificate margin & 0.002176 \\
Largest observed $|\ell_{s,a}|$ against $B=0.05$, reference path & 0.0081 \\
Openings influenced by a same-date maturity & 0 \\
Maximum split-account aggregation identity error & $2\times10^{-16}$ \\
\bottomrule
\end{tabular}
\end{center}
\end{table}

\begin{table}[t]
\caption{Minimum daily protected wealth relative to the induced \Safe{} benchmark. The deterministic floor is $\rho=0.80$ for every path. The first, third, and fourth rows are reproduced at full precision by the deposited result tables. The feature-only row is the pooled feature-only Ridge transferred from the 24-stock universe described in the manuscript's Experimental Design; only its saved run summary survives in the archive, and it is reported as archived evidence.}
\label{tab:protected_paths}
\begin{center}
\small
\begin{tabular}{lrr}
\toprule
Path & Minimum $W_t^{\mathrm{prot}}/S_t$ & Margin above $\rho$ \\
\midrule
Primary signed path & 0.998406 & 0.198406 \\
Feature-only 24-to-82 transfer (archived summary) & 0.998995 & 0.198995 \\
Legacy full raw diagnostic & 0.999722 & 0.199722 \\
Clean feature-component diagnostic & 0.998936 & 0.198936 \\
\bottomrule
\end{tabular}
\end{center}
\end{table}

The three tables together show every audited invariant holding with margin on the realized path. Feasibility never required contraction, no intervention ever created exposure and every removal landed on \Safe{} exactly, and all four audited daily paths stayed far above the deterministic floor, while the induced \Safe{} benchmark itself dipped to roughly $0.942$, so the protection measured here is against incremental prediction-assisted damage, not against market risk in the fallback. The manuscript's certificates describe the code that actually ran.

\section{Layer ablation and synthetic regime placement}\label{s:ablation}

The manuscript's ablation table reports the primary signal. Table~\ref{tab:ablation_all} gives the same seven variants on all three real signal panels, each run through the identical split account with only the named layers disabled, measured against the induced \Safe{} benchmark. Table~\ref{tab:synthetic_regimes} gives the regime-level placement behind the manuscript's synthetic ladder, means over ten seeds of the complete controller's executed \Follow{} and \Reverse{} mass, its abstention fraction, and the realized mean rewards of the two orientations within each regime.

\begin{table}[t]
\caption{Layer ablation on the three real signal panels. Terminal gain and worst shortfall are in basis points of wealth relative to the induced \Safe{} benchmark. Always \Follow{} and Always \Reverse{} are fixed-action references through the same split account.}
\label{tab:ablation_all}
\begin{center}
\footnotesize
\setlength{\tabcolsep}{4pt}
\begin{tabular}{llrrrr}
\toprule
Signal & Variant & Terminal gain & Worst shortfall & Mean active prob. & Abstention \\
\midrule
Clean full raw & Full controller & $-15.9$ & 15.9 & 0.108 & 62.0\% \\
 & No gate & $-10.6$ & 10.6 & 0.250 & 0.0\% \\
 & No budget & $-16.5$ & 16.5 & 0.122 & 61.8\% \\
 & Neither & $-26.8$ & 26.8 & 0.631 & 0.0\% \\
 & Always \Follow{} & $+13.4$ & 18.2 & 1.000 & 0.0\% \\
 & Always \Reverse{} & $-45.9$ & 57.3 & 1.000 & 0.0\% \\
\midrule
Legacy full raw & Full controller & $-2.8$ & 2.8 & 0.117 & 61.1\% \\
 & No gate & $-10.6$ & 10.6 & 0.250 & 0.0\% \\
 & No budget & $-1.7$ & 1.7 & 0.125 & 60.9\% \\
 & Neither & $-26.9$ & 26.9 & 0.631 & 0.0\% \\
 & Always \Follow{} & $-5.2$ & 21.4 & 1.000 & 0.0\% \\
 & Always \Reverse{} & $-25.1$ & 38.9 & 1.000 & 0.0\% \\
\midrule
Clean feature component & Full controller & $-9.7$ & 10.6 & 0.095 & 64.8\% \\
 & No gate & $-10.6$ & 10.6 & 0.249 & 0.0\% \\
 & No budget & $-10.3$ & 11.3 & 0.113 & 64.5\% \\
 & Neither & $-26.9$ & 26.9 & 0.631 & 0.0\% \\
 & Always \Follow{} & $+17.6$ & 47.2 & 1.000 & 0.0\% \\
 & Always \Reverse{} & $-54.5$ & 61.3 & 1.000 & 0.0\% \\
\bottomrule
\end{tabular}
\end{center}
\end{table}

\begin{table}[t]
\caption{Synthetic regime placement of the complete controller, means over ten seeds. Reward means are in basis points per episode.}
\label{tab:synthetic_regimes}
\begin{center}
\footnotesize
\setlength{\tabcolsep}{4pt}
\begin{tabular}{llrrrrr}
\toprule
Strength & Regime & \Follow{} mass & \Reverse{} mass & Abstention & \Follow{} reward & \Reverse{} reward \\
\midrule
0.00 & \Follow{} & 0.006 & 0.001 & 97.4\% & $-0.47$ & $-0.47$ \\
 & neutral & 0.005 & 0.001 & 98.0\% & $-0.40$ & $-0.54$ \\
 & \Reverse{} & 0.005 & 0.002 & 97.6\% & $-0.41$ & $-0.53$ \\
0.03 & \Follow{} & 0.257 & 0.000 & 9.7\% & $+4.61$ & $-5.44$ \\
 & neutral & 0.040 & 0.001 & 87.3\% & $-0.39$ & $-0.53$ \\
 & \Reverse{} & 0.000 & 0.293 & 9.1\% & $-5.22$ & $+4.41$ \\
0.06 & \Follow{} & 0.277 & 0.000 & 9.7\% & $+9.63$ & $-10.45$ \\
 & neutral & 0.052 & 0.001 & 84.4\% & $-0.39$ & $-0.53$ \\
 & \Reverse{} & 0.000 & 0.307 & 5.6\% & $-10.09$ & $+9.29$ \\
0.12 & \Follow{} & 0.301 & 0.000 & 9.7\% & $+19.67$ & $-20.48$ \\
 & neutral & 0.064 & 0.001 & 82.5\% & $-0.38$ & $-0.52$ \\
 & \Reverse{} & 0.000 & 0.320 & 3.8\% & $-19.84$ & $+19.05$ \\
\bottomrule
\end{tabular}
\end{center}
\end{table}

The ablation ordering the manuscript reports for the primary signal, no gate below the full controller in shortfall, holds for the clean feature component as well, where the two are equal, and reverses for the legacy signal, where the full controller's 2.8 basis points sits below the no-gate variant's 10.6 and the no-budget variant does better still. The budget's cap binds identically across signals, which is why every no-gate row shows the same 0.250 mean active probability and the same 10.6 shortfall, whereas the gate's effect depends on which orientation the matured record supports on each panel. The regime table shows the abstention of the synthetic ladder concentrating where it should. In neutral regimes the controller withholds 82\% to 87\% of decisions at every positive strength, and in \Follow{} and \Reverse{} regimes it places its executed mass almost entirely on the orientation whose realized reward is positive, with the opposite orientation's mass below 0.001. What these tables do not establish is market evidence of engagement, because no real signal in the archive was informative enough to produce it.

\fi

\end{document}